\documentclass[journal]{IEEEtran}

\usepackage{setspace}

\usepackage{amsmath, amsfonts}

\usepackage{cite}
\usepackage{algorithm}
\usepackage{algpseudocode}
\usepackage{enumerate}
\usepackage{color}
\usepackage{balance}
\usepackage{url}
\usepackage{multirow}  
\usepackage{pbox}  
\usepackage{flushend} 

\usepackage[pdftex]{graphicx}
\graphicspath{{./}}%
\DeclareGraphicsExtensions{.jpg,.png,.pdf}
\usepackage[font={footnotesize}]{caption}
\usepackage[caption=false,font=footnotesize]{subfig}

\newcommand{\bi}{\begin{itemize}}	
\newcommand{\ei}{\end{itemize}}
\newcommand{\bn}{\begin{enumerate}}	
\newcommand{\en}{\end{enumerate}}
\newcommand{\bc}{\begin{center}}
\newcommand{\ec}{\end{center}}
\newcommand{\be}{\begin{equation}}
\newcommand{\ee}{\end{equation}}
\newcommand{\bea}{\begin{eqnarray}}
\newcommand{\eea}{\end{eqnarray}}
\newcommand{\ben}{\begin{equation*}}
\newcommand{\een}{\end{equation*}}
\newcommand{\beqa}{\begin{eqnarray}}
\newcommand{\eeqa}{\end{eqnarray}}

\newcommand{\red}{\textcolor{red}} 
\newcommand{\blue}{\textcolor{blue}}

\begin{document}
\title{Measurement Analysis and Channel Modeling for TOA-Based Ranging in Tunnels}

\author{Vladimir Savic, Javier Ferrer-Coll, Per \"{A}ngskog, Jos\'{e} Chilo, \\Peter Stenumgaard, and Erik G. Larsson
\thanks{Copyright (c) 2014 IEEE. Personal use of this material is permitted. However, permission to use this material for any other purposes must be obtained from the IEEE by sending a request to pubs-permissions@ieee.org. The original version of this manuscript is published in IEEE Transactions on Wireless Communications (DOI: 10.1109/TWC.2014.2350493).}
\thanks{V. Savic and E. G. Larsson are with the Dept. of Electrical Engineering (ISY), Link\"{o}ping University, Sweden (e-mails: vladimir.savic@liu.se, erik.larsson@isy.liu.se). J. Ferrer-Coll, P. \"{A}ngskog, and J. Chilo are with the Dept. of Electronics, Mathematics and Natural Sciences, University of G\"{a}vle, Sweden (e-mails: Javier.FerrerColl@hig.se, Per.Angskog@hig.se, Jose.Chilo@hig.se). P. Stenumgaard is with the Dept. of Electrical Engineering (ISY), Link\"{o}ping University, Sweden, and the Swedish Defense Research Agency (FOI) (e-mail: peterst@isy.liu.se).}%
\thanks{The work was supported by the project Cooperative Localization (CoopLoc), funded by the Swedish Foundation for Strategic Research (SSF).}
}

\maketitle

\begin{abstract}
A robust and accurate positioning solution is required to increase the safety in GPS-denied environments. Although there is a lot of available research in this area, little has been done for confined environments such as tunnels. Therefore, we organized a measurement campaign in a basement tunnel of Link\"{o}ping university, in which we obtained ultra-wideband (UWB) complex impulse responses for line-of-sight (LOS), and three non-LOS (NLOS) scenarios. This paper is focused on time-of-arrival (TOA) ranging since this technique can provide the most accurate range estimates, which are required for range-based positioning. We describe the measurement setup and procedure, select the threshold for TOA estimation, analyze the channel propagation parameters obtained from the power delay profile (PDP), and provide statistical model for ranging. According to our results, the rise-time should be used for NLOS identification, and the maximum excess delay should be used for NLOS error mitigation. However, the NLOS condition cannot be perfectly determined, so the distance likelihood has to be represented in a Gaussian mixture form. We also compared these results with measurements from a mine tunnel, and found a similar behavior.
\end{abstract}
\begin{keywords}
tunnels, channel modeling, time of arrival, ultra-wideband, impulse response, ranging, positioning.
\end{keywords}

\section{Introduction}\label{sec:intro}

Accurate positioning can enable many applications \cite{Sayed2005} including positioning of employees and rescue personnel in industrial environments. For example, knowledge of the last location of the miner, in the aftermath of a mine collapse or explosion, is crucial for search-and-rescue operations \cite{Novak2009}. Although there is a lot of available research in this area, little has been done for confined environments such as tunnels. Exceptions are few a proposals based on trilateration \cite{Chehri2009} and fingerprinting \cite{Nerguizian2006} techniques, but they are not sufficiently robust against outliers and changes in the environment. Therefore, more research is required to provide accurate channel models, especially for range-based positioning algorithms.

In this paper, we present the result of a measurement campaign in a basement tunnel of Link\"{o}ping university in Sweden (referred to as the LiU tunnel). Motivated by high temporal resolution of the signals with large bandwidth \cite{Gezici2005}, we decided to use UWB signal. We obtained UWB complex impulse responses for LOS, and three NLOS scenarios, in which the direct path is blocked by a metal obstacle, a person and a tunnel wall. Then, we focus on TOA-based ranging since this technique can provide the most accurate range estimates, which are required for range-based positioning. More specifically, we i) describe the measurement setup and procedure, ii) determine the threshold for TOA estimation using trade-off between false alarms and missed detections, iii) analyze channel propagation parameters obtained from PDPs, and iv) provide statistical model for ranging. According to our results, the rise-time should be used for NLOS identification, and the maximum excess delay should be used for NLOS error mitigation. However, the NLOS condition cannot be perfectly determined due to the overlap in all considered parameters, so the distance likelihood has to be represented in a Gaussian mixture form. Our main contribution is a statistical model that is especially suitable for range-based Bayesian positioning or tracking algorithms, but which can be also used for many other deterministic algorithms. We also compared these results with the measurements from an iron-ore mine tunnel (located in Kiruna, Sweden), and found that our main conclusions are valid also for this measurement set.

The remainder of this paper is organized as follows. In Section \ref{sec:rel}, we review other indoor measurement campaigns, and NLOS identification and error mitigation techniques. In Section \ref{sec:setup}, we describe our experimental setup and measurement procedure. Then, in Section \ref{sec:ch-param}, we define channel propagation parameters, and select the threshold for TOA estimation. Measurement analysis and channel modeling for TOA-based ranging are performed in Section \ref{sec:analysis-model}. A comparison with measurements from the mine tunnel is shown in Section \ref{sec:kiruna}. Finally, Section \ref{sec:conc} provides our conclusions and proposals for future work.

\section{Related work}\label{sec:rel}

\subsection{Overview of measurements campaigns}\label{subsec:campaigns} 
Multipath propagation has been studied and characterized in a multitude of environments, such as residential buildings, indoor offices and mines. The work performed in \cite{Ghassemzadeh2004a} proposes statistical models for estimating the root-mean-square (RMS) delay spread, path-loss and other relevant propagation parameters. By modeling all parameters as random variables, their models are capable to predict propagation in homes in which the measurements are not performed. In \cite{Ghassemzadeh2005a}, they also provide  models for the UWB power-delay profile (PDP) in residential areas. The work in \cite{Kivinen2001} contains a wideband channel characterization of indoor office environments, in which medium levels of RMS delay spread and low levels of path-loss are observed. Similar levels of delay spread are found for UWB channels in an indoor lab environment in \cite{Denis2003}. All these environments \cite{Ghassemzadeh2004a,Ghassemzadeh2005a,Kivinen2001,Denis2003} have delay spread values not higher than 50 ns. 
On the other hand, industrial environments are characterized as reflective environments with high delay spread values. However, the study performed in \cite{Ferrer-Coll2012c} distinguished industrial environments with opposite propagation characteristics, i.e., a steel mill and a paper mill, with 298 ns and 23 ns RMS delay spread, respectively. 

\begin{table}[!tb]
\caption{Summary of measurements campaigns in different environments with main parameters ($f$ - frequency, $B$ - bandwidth, and $\tau_{RMS}$ - RMS delay spread.). Other relevant parameters can be found in the cited publications.  }
\centering
\begin{tabular}{l||c|c|c}
\multirow{4}{*} ~Environment & $f$ [GHz] & $B$ [GHz] & $\tau_{RMS}$ [ns] \\ \hline\hline
Residence   \cite{Ghassemzadeh2005a} & 5  & 6   & \pbox{20cm}{4.7-8.2} \\ \hline 
Indoor office \cite{Kivinen2001} & 5.3  & 0.053 &  \pbox{20cm}{30-50} \\ \hline
Indoor lab \cite{Denis2003} & 4  & 4 &  \pbox{20cm}{8-20} \\ \hline  
Commercial \cite{Ghassemzadeh2005a} & 5  & 6  &  \pbox{20cm}{5.5-8.2}  \\ \hline
Steel mill \cite{Ferrer-Coll2012c} & 1.89  & 0.5  & 298  \\ \hline
Paper mill  \cite{Ferrer-Coll2012c} & 1.89  & 0.5 & 23  \\ \hline
Subway tunnel \cite{He2013} & 2.4  & 0.1 &  \pbox{20cm}{159-234}   \\ \hline 
Road tunnel \cite{Ching2009} & 5.2  &  0.1 & 20-100   \\ \hline 
Mine tunnel \cite{Sun2010} & 1  & 0.2 & 20-50\\ \hline 
Mine tunnel \cite{Ferrer-Coll2012b} & 2.4   & 0.5 & 14\\ \hline 
Mine tunnel \cite{Nerguizian2005} & 2.4  & 0.2 & 27.4 \\ \hline
Mine tunnel \cite{Boutin2008} & 2.4/5.8  & 0.2 & 1-15 \\ \hline
Mine tunnel \cite{Chehri2012} & 3.5  & 3 & 11-29 \\ \hline
Mine tunnel \cite{Mabrouk2012} & 2.4  & 0.2 & 1.7 \\ \hline
LiU tunnel & 3.5  & 2 & 3-16\\
\end{tabular}
\label{Table_Parameters}
\end{table}

Regarding previous measurements in tunnels, the work in \cite{He2013} presents measurements in a subway tunnel, where it is found that delay spread levels can go up to 234 ns due to multiple reflections in the subway station. Analysis of measurements obtained in an arched tunnel \cite{Ching2009} showed a waveguide effect, as expected for tunnel-like environments. In this work, the authors separate the multipath components from the different surfaces, and found that scattering from the ground was dominant. In \cite{Sun2010}, the authors proposed a multi-mode model for predicting the received power and the PDP in tunnels with pillars. The analysis of measurements performed in wide and narrow iron-mine tunnels \cite{Ferrer-Coll2012b} show a small delay spread even in NLOS scenarios. The work in \cite{Nerguizian2005} analyzed the channel of an underground gold mine, and found that the RMS delay spread is decreasing with distance. Moreover, the same environment is considered in \cite{Boutin2008,Chehri2012}, where the authors found that the RMS delay spread is almost uncorrelated with the distance, and in \cite{Mabrouk2012}, where the authors found that the RMS delay spread can be significantly reduced ($<2$ ns) using directional antennas. Generally, the studies performed in mine tunnels have found low delay spread levels due to the particular structure of the environment. The multipath rays reflect against  the walls, the ceiling and the floor between the transmitter and the receiver, but not from the back of the transmitter and the receiver. Moreover, in contrast to indoor office environments, the delay spread is typically not increasing with distance. 

A summary of these measurement campaigns, including LiU tunnel (to be discussed in next sections), is shown in Table~\ref{Table_Parameters}. For a more detailed overview of measurement campaigns and channel models, we refer the reader to \cite{Molisch2003,Molisch2004,Forooshani2013}.

\subsection{Overview of NLOS identification and error mitigation techniques}\label{sec:nlos-overview}
TOA-based ranging based on NLOS measurements typically leads to positively biased estimates. There are many proposals in literature for how to deal with this problem, which can be broadly classified into two categories \cite{Khodjaev2010}: i) NLOS identification, and ii) NLOS error mitigation. The former one attempts to distinguish between LOS and NLOS conditions, while the latter one attempts to reduce the bias caused by an NLOS condition assuming that this NLOS condition is identified.

NLOS identification can be performed by analyzing the variance of the time-series of the range estimates \cite{Wylie1996}. Since the NLOS measurements typically have much larger variance, the hypothesis testing can be easily performed. However, this approach would lead to high latency since it requires a large number of measurements. An alternative approach is to use channel propagation parameters from the complex impulse response. For instance, in \cite{Venkatesh2007a} three parameters are jointly used (RMS delay-spread, TOA and RSS) to distinguish between LOS and NLOS scenarios. They found that RMS delay spread is the most useful for this problem, but the combination of these three parameters can improve the probability of correct identification. In \cite{Zhang2013}, the authors found that the kurtosis provides consistent information about NLOS condition, and that using multiple antennas can improve this information. In \cite{Marano2010}, multiple parameters are considered by using a nonparametric least-square support-vector-machine (LS-SVM) classifier. This approach does not require statistical models, since it directly works with training samples. A nonparametric approach is also used in \cite{Gezici2003}, where the authors use the training samples to construct the kernel of the LOS and NLOS error probability density functions (PDFs). Then, they use Kullback-Leibler (KL) divergence to measure the distance between these PDFs, and set the decision threshold.

Once NLOS identification is performed, the measurement can be discarded but it would lead to unnecessary loss of useful information (especially, if there are no sufficient LOS links). Therefore, NLOS error mitigation is required to make NLOS measurements useful for ranging. Since the distribution of the NLOS error depends on the spatial distribution of the scatterers, the mitigation could be performed by modeling these scatterers \cite{Al-Jazzar2002}. However, this approach is   typically not feasible due to the complex shape of the environment, and possible dynamic obstacles. Another way is to model the NLOS error as a function of some channel propagation parameter. For example, in \cite{Denis2003}, the authors found that the NLOS error is increasing with the mean excess delay and the RMS delay spread. Therefore, a simple polynomial model can be used to significantly reduce this error. Nonparametric regression can be also used to compute the NLOS error as a function of multiple channel propagation parameters. For this purpose, LS-SVM regression has been used in \cite{Marano2010}, and Gaussian process regression in \cite{Wymeersch2012a}. Finally, in some cases, it may not be possible to detect an NLOS condition, but only its probability. In that case, a soft-decision approach is required, in which NLOS identification and error mitigation are combined into one single step. This approach is proposed in \cite{Cong04}, in which the ranging distribution is a mixture of LOS and NLOS models. In addition, this work proposes to use three different models for NLOS errors, depending how much a priori information is available.

A detailed survey of NLOS identification and error mitigation techniques can be found in \cite{Khodjaev2010}.

\section{Experimental setup and measurement procedure}\label{sec:setup}

\begin{figure}[!tb]
\centerline{
\subfloat[]{\includegraphics[width=0.48\columnwidth]{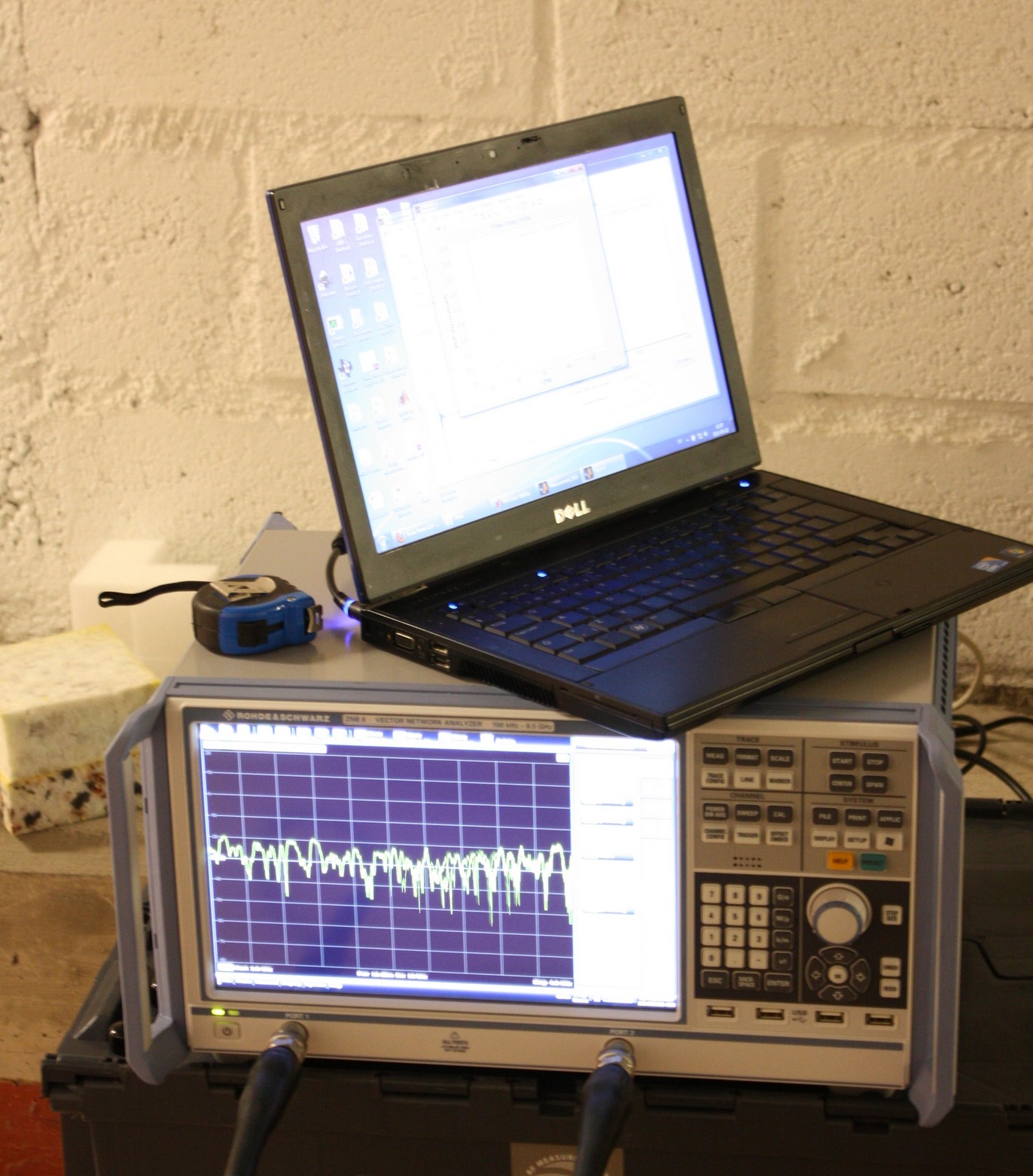}\label{fig:vna-pc}}
\hfill
\subfloat[]{\includegraphics[width=0.455\columnwidth]{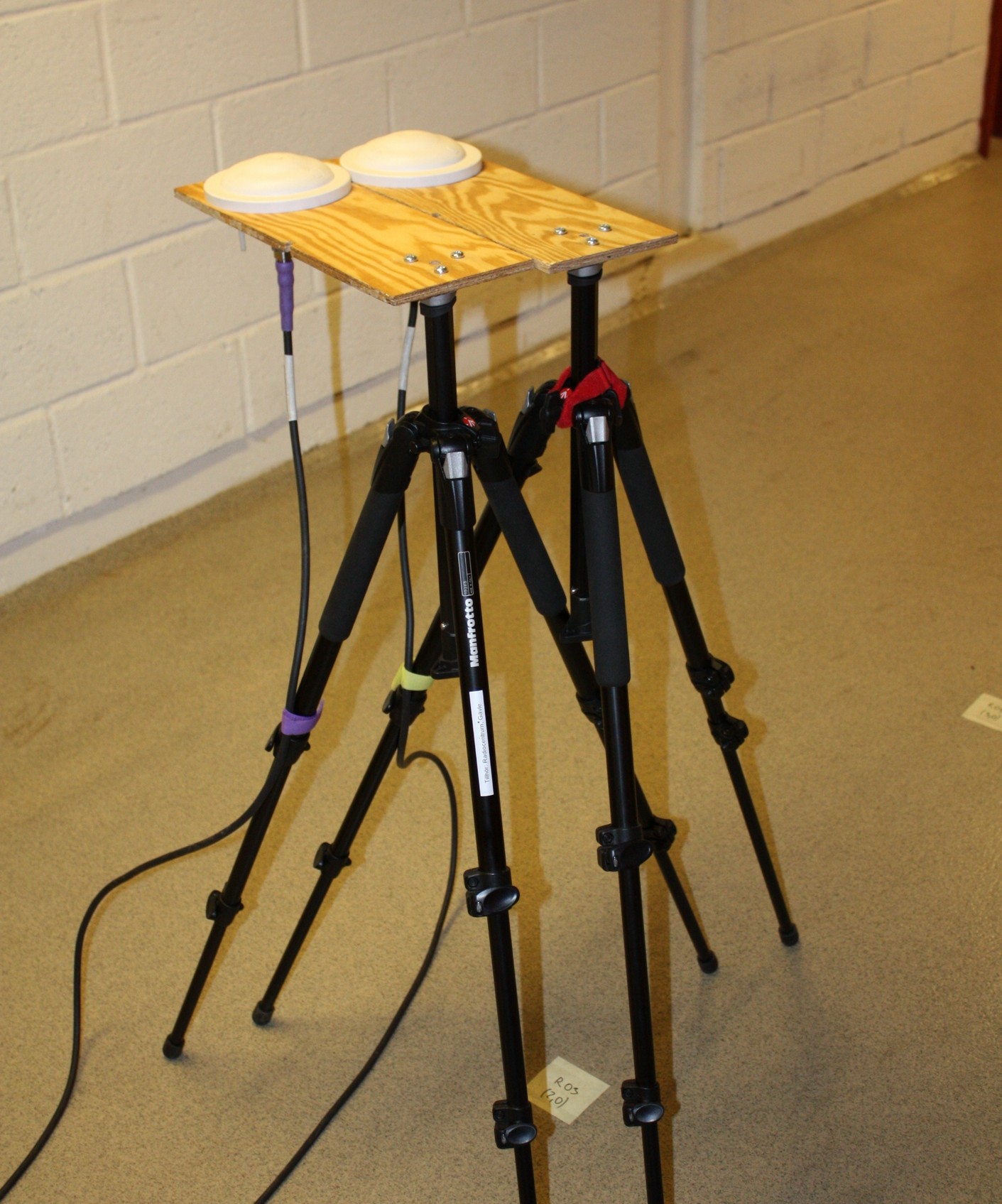}\label{fig:antennas}}
}
\caption{Experimental setup: (a) VNA connected to PC, and (b) omni-directional UWB antennas.}
\label{fig:equip}
\end{figure}

\begin{table}[!tb]
\caption{Measurement parameters}\label{table:param}
\centering
\begin{tabular}{l l}
\hline\hline
Signal power & 12 dBm \\
Waveform & sinusoidal sweep\\
Center frequency & 3.5 GHz\\
Bandwidth & 2 GHz\\
IF filter & 10 KHz\\
Number of points & 3001\\
Sweep time & 263 ms\\
Time resolution & 0.5 ns\\
Antenna range & 1.71 - 6.4 GHz\\
Antenna gain & 5 - 7.5 dBi\\
Cable attenuation & 0.65 dB/m\\
Bandpass filter & Hann window\\
\hline \hline
\end{tabular}
\end{table} 

The measurement setup (Fig. \ref{fig:equip}) consists of a vector network analyzer (VNA), two ultra-wide band (UWB) omni-directional antennas and coaxial cables to connect the antennas with the VNA. A PC is used to set the VNA parameters and extract the multiple frequency responses from the instrument. In our case, we use a swept-frequency sinusoidal  signal (with 3001 points) to characterize the channel between 2.5 and 4.5 GHz. The power level was set to 12 dBm, and a calibration of the system is performed to compensate for the effect of VNA, cables and antennas (i.e., received power was shifted to 0 dBm when transmitter and receiver were placed as in Fig. \ref{fig:equip}b). Then, the frequency responses are transferred to the PC where a Hann window \cite{Sanchez2001} is used to reduce the out-of-band noise, and to ensure causality of the time-domain responses. Finally, by applying the inverse fast Fourier transform, the complex impulse responses are estimated, and subsequently, PDPs are calculated. We summarize the parameters in Table \ref{table:param}.

\begin{figure}[!tb]
\centerline{
\subfloat[]{\includegraphics[width=0.48\columnwidth]{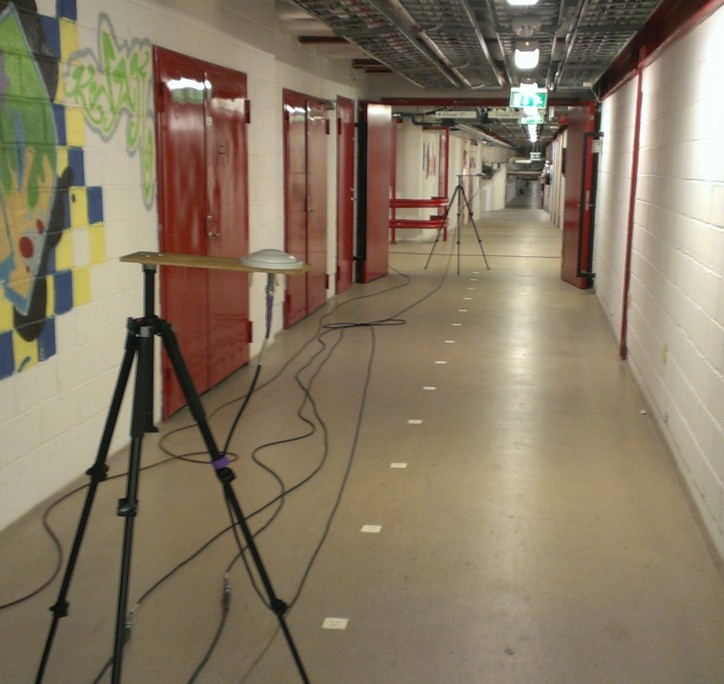}\label{fig:tunnel1-los}}
\hfill
\subfloat[]{\includegraphics[width=0.48\columnwidth]{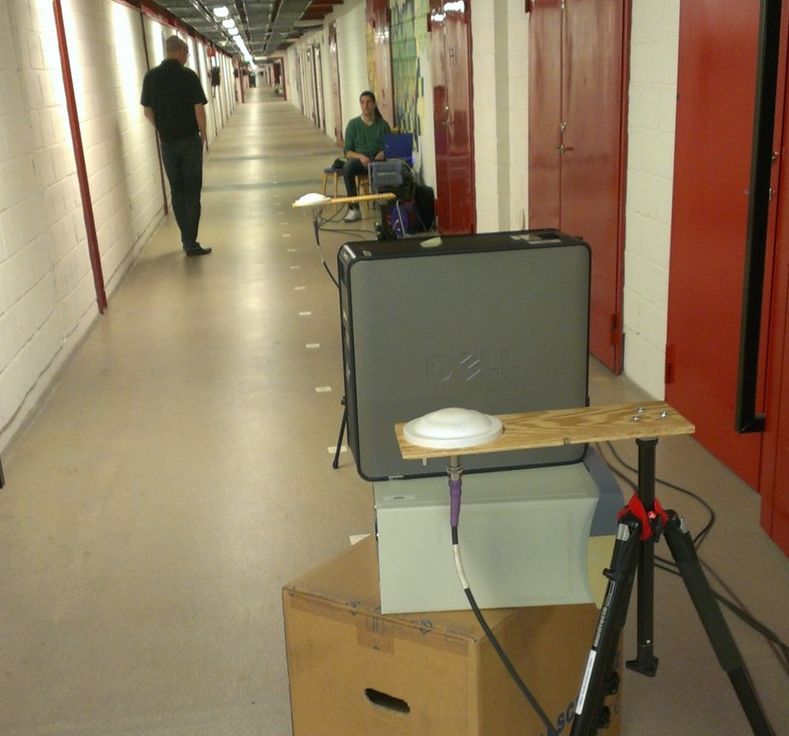}\label{fig:tunnel1-nlosm}}
}
\centerline{
\subfloat[]{\includegraphics[width=0.48\columnwidth]{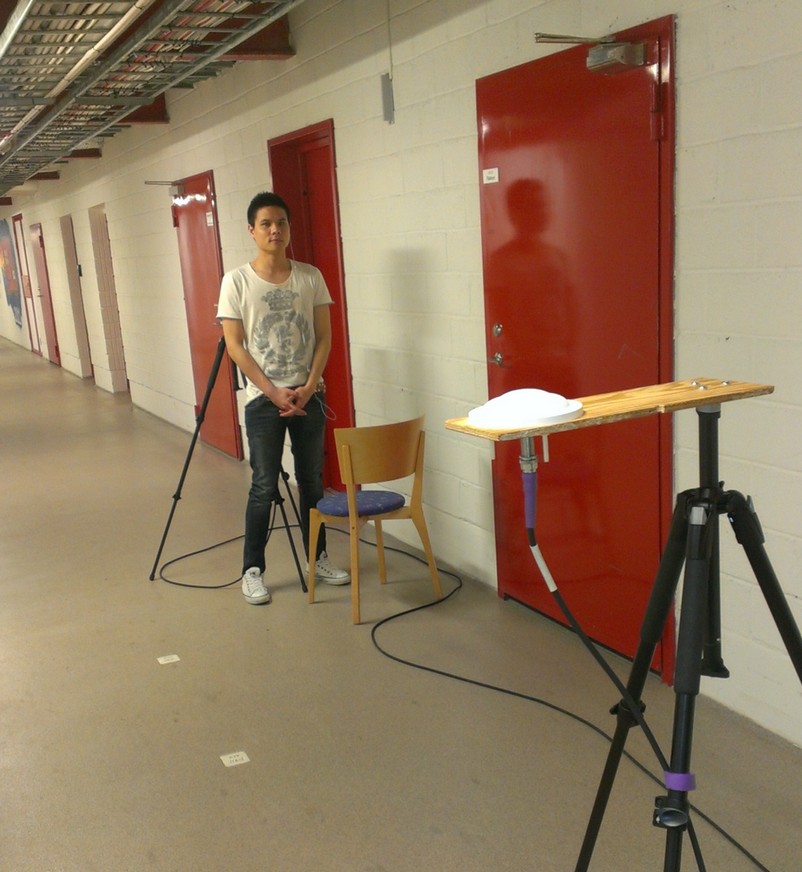}\label{fig:tunnel1-nlosp}}
\hfill
\subfloat[]{\includegraphics[width=0.48\columnwidth]{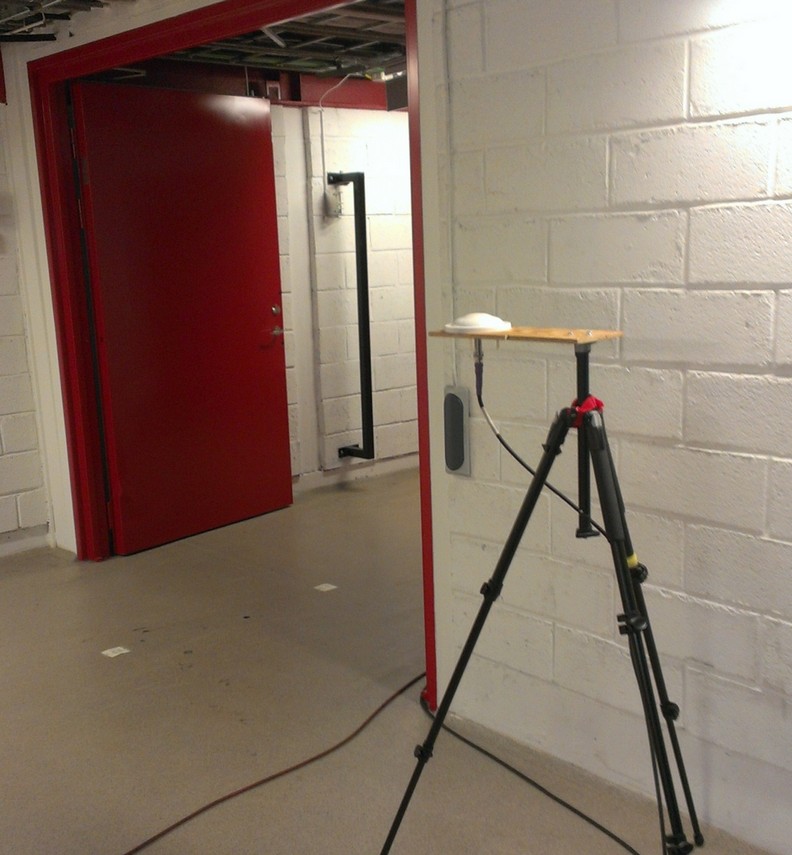}\label{fig:tunnel1-nlosw}}
}
\caption{Considered scenarios: (a) LOS, (b) NLOS-M, (c) NLOS-P, and (d) NLOS-W. The obstacles were placed in front of the transmitter antenna.}
\label{fig:scenarios}
\end{figure}

\begin{figure*}[!tb]
\centerline{
\includegraphics[width=0.85\textwidth]{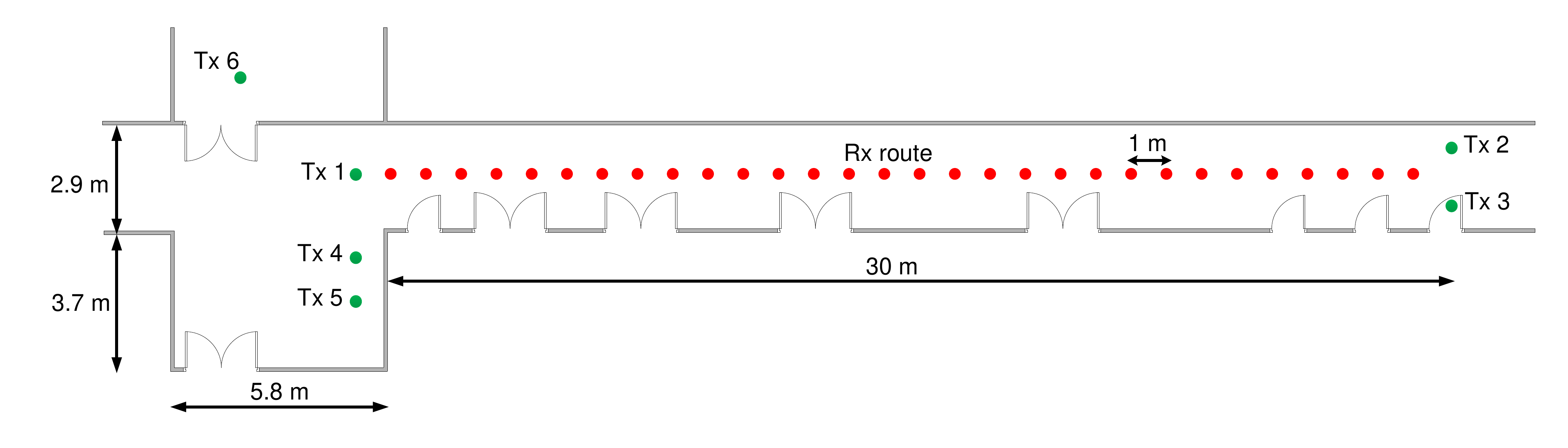}}
\caption{Deployment of transmitters (Tx) and receivers (Rx) in the LiU tunnel. There are 6 transmitter positions (marked with green circles), and 30 receiver positions (marked with red circles). All doors were closed, except for the one next to the position of  Tx 6. The height of the tunnel is 2.8 m.}
\label{fig:tunnel-deploy}
\end{figure*}

Using the described setup, we performed the measurements in a basement tunnel of LiU (Fig. \ref{fig:scenarios}), which has similar characteristics to tunnels in industrial environments. The tunnel walls, excluding the metal doors, are built of concrete blocks with steel reinforcement. The ceiling is also made of concrete, but with many metal pipes.
We considered four different scenarios: LOS, and NLOS caused by three different obstacles: a metal sheet, a person, and a tunnel wall (denoted by NLOS-M, NLOS-P, and NLOS-W, respectively). For each of these scenarios, we placed the transmitter in 3 locations and receiver in 30 locations forming the route through the tunnel. For each transmitter-receiver pair, we obtained 10 PDPs, so we obtained 3600 PDPs in total (900 per scenario). All considered scenarios are shown in Fig. \ref{fig:scenarios}, and the deployment is shown in Fig. \ref{fig:tunnel-deploy}.

\section{Channel propagation parameters and threshold selection}\label{sec:ch-param}

Given complex impulse responses of the channel, $h(t) = \sum\nolimits_{k = 1...{N_k}} {{a_k}\delta (t - {\tau _k})}$ ($N_k=3001$, $a_k \in \mathbb{C}$), we can obtain the PDP as $\left|h(t)\right|^2$ \cite{Nerguizian2005}. However, since most of the components of the PDP are caused by thermal noise, we consider only components above a certain threshold ${P_{TH}}$ [dBm], i.e.,
\be\label{eq:pdp-thresh}
{p_h}(t) = \left\{ \begin{array}{l}
{\left| {h(t)} \right|^2},\,{\rm{if}}\,\,10\log_{10} \left({\left| {h(t)} \right|^2}/P_0\right) > {P_{TH}}\\
0,\,\,\,\,\,\,\,\,\,\,\,\,\,{\rm{otherwise}}
\end{array} \right.
\ee
where $P_0=1$ mW. Then, we consider following channel propagation parameters:
\bi
\item \textit{Time of arrival (TOA)}: 
\be\label{eq:toa}
{\tau _1} = \min \{ t:\,{p_h}(t) > 0\}
\ee
\item \textit{Received signal strength (RSS)} [dBm]: 
\be\label{eq:rss}
{P_{\rm{RSS}}} = 10\log_{10} \left( {\frac{1}{T P_0} \int\limits_0^T {{p_h}(t)dt}} \right) 
\ee
where $T=1.5~\mu s$ is the observation interval.
\item \textit{Maximum received power} [dBm]:
\be\label{eq:mrp}
{P_{{\rm{MAX}}}} = 10\log_{10} \left( \frac{1}{P_0}{\max_t {{p_h}(t)}} \right)
\ee
\item \textit{Mean excess delay}:
\be\label{eq:med}
\bar \tau  = \frac{{\int\limits_0^T {t{p_h}(t)dt} }}{{\int\limits_0^T {{p_h}(t)dt} }}
\ee
\item \textit{Maximum excess delay}:
\be\label{eq:mxed}
{\tau _{{\rm{MAX}}}} = \max \{ t:\,{p_h}(t) > 0\}  - {\tau _1}
\ee
which is a measure of \textit{total} delay spread of the PDP.
\item \textit{Root-mean-square (RMS) delay spread}:
\be\label{eq:rds}
{\tau _{{\rm{RMS}}}} = \sqrt{\frac{{\int\limits_0^T {{{(t - \bar \tau )}^2}{p_h}(t)dt} }}{{\int\limits_0^T {{p_h}(t)dt} }}}
\ee
which is a measure of \textit{effective} delay spread of the PDP.
\item \textit{Rise time}:
\be\label{eq:rt}
\tau_{{\rm{RT}}}=\arg\max_t {{p_h}(t)}-\tau_{1}
\ee
\item \textit{Kurtosis}:
\be\label{eq:kurt}
\kappa  = \frac{{\frac{1}{T}\int\limits_0^T {{{\left( {\left| {h(t)} \right| - {\mu _{\left| h \right|}}} \right)}^4}dt} }}{{{{\left( {\frac{1}{T}\int\limits_0^T {{{\left( {\left| {h(t)} \right| - {\mu _{\left| h \right|}}} \right)}^2}dt} } \right)}^2}}}
\ee
where ${\mu _{\left| h \right|}} = \frac{1}{T}\int\limits_0^T {\left| {h(t)} \right|dt}$. Kurtosis is dimensionless metric that quantifies how $\left| {h(t)} \right|$ matches the Gaussian distribution (larger $\kappa$ implies stronger non-Gaussianity).
\ei

\begin{figure}[!tb]
\centerline{
\subfloat[]{\includegraphics[width=0.51\columnwidth]{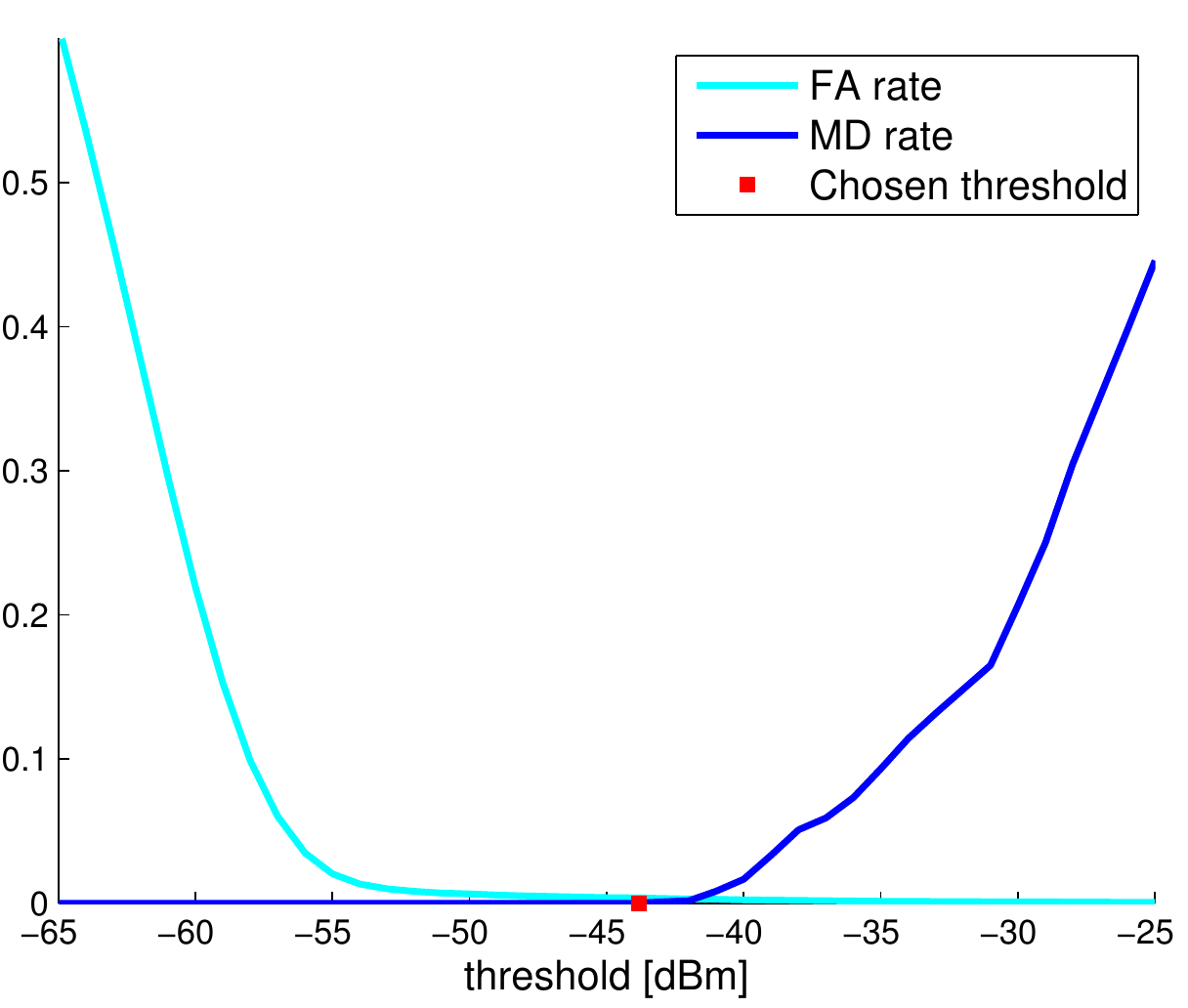}\label{fig:fa-md-rate}}
\hfill
\subfloat[]{\includegraphics[width=0.51\columnwidth]{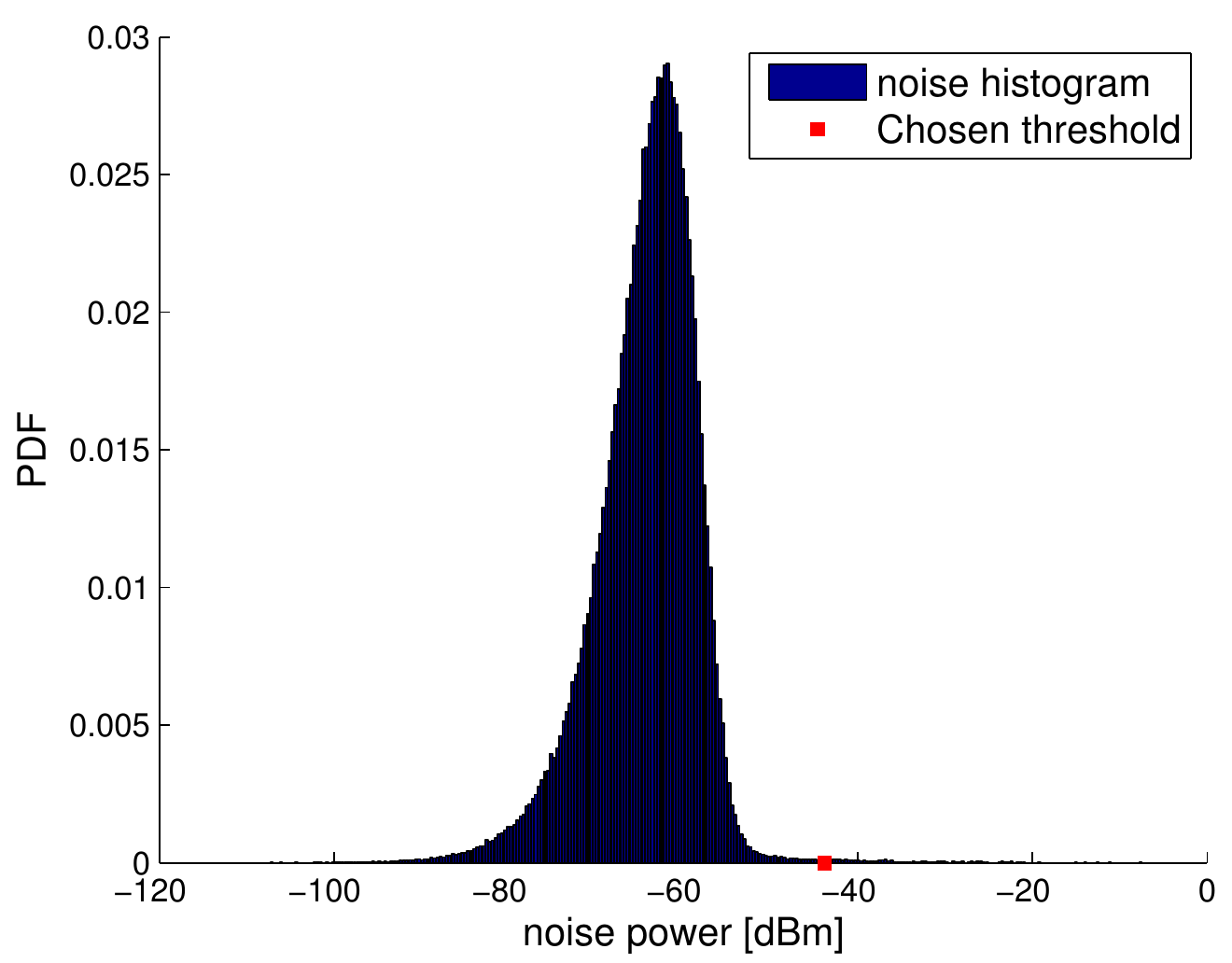}\label{fig:noise-hist}}
}
\caption{(a) Chosen threshold based on trade-off between FA and MD rate, and (b) PDF of the thermal noise. The power levels are calibrated.}
\label{fig:thresh-sel}
\end{figure}

Then, we need to choose the threshold $P_{TH}$. There are many ways to do it \cite{Dardari2009}, but since our main goal is robust TOA-based ranging, we decide to choose the value which provides a good trade-off between false-alarm (FA) (when noise is detected instead of the signal) and missed-detection (MD) (when threshold is higher than the strongest path) rates. Therefore, we obtain the FA and MD rate for all reasonable values of the threshold\footnote{We work with calibrated power level since the real power level is not needed for our problem.}. Taking into account the results in Fig. \ref{fig:thresh-sel}a, we set $P_{TH}=-43.8$ dBm. We can see that, according to the thermal noise PDF (Fig. \ref{fig:thresh-sel}b), this value of the threshold is on very right tail of the PDF (i.e, ${P_{TH}} = {\mu_{noise}} + 3.4{\sigma _{noise}} =  - 43.8\,{\rm{dBm}}$, where ${\mu_{noise}} =  - 64\,{\rm{dBm}}\,,\,{\sigma _{noise}} = 6\,\,{\rm{dB}}$). Note that the chosen threshold does not minimize the root-mean-square-error (RMSE) in the TOA estimation ($P_{TH}=-51$ dBm would achieve it, but it would lead to many false alarms). In principle, this is not a problem since we will in addition perform error mitigation for NLOS range estimates, as will be shown in Section \ref{subsec:stat-models}. Note also that the threshold is a function of the SNR in the general case (i.e., the transmit power and maximum range), so it should be adapted to the considered scenario.

\begin{figure}[!tb]
\centerline{
\subfloat[]{\includegraphics[width=0.51\columnwidth]{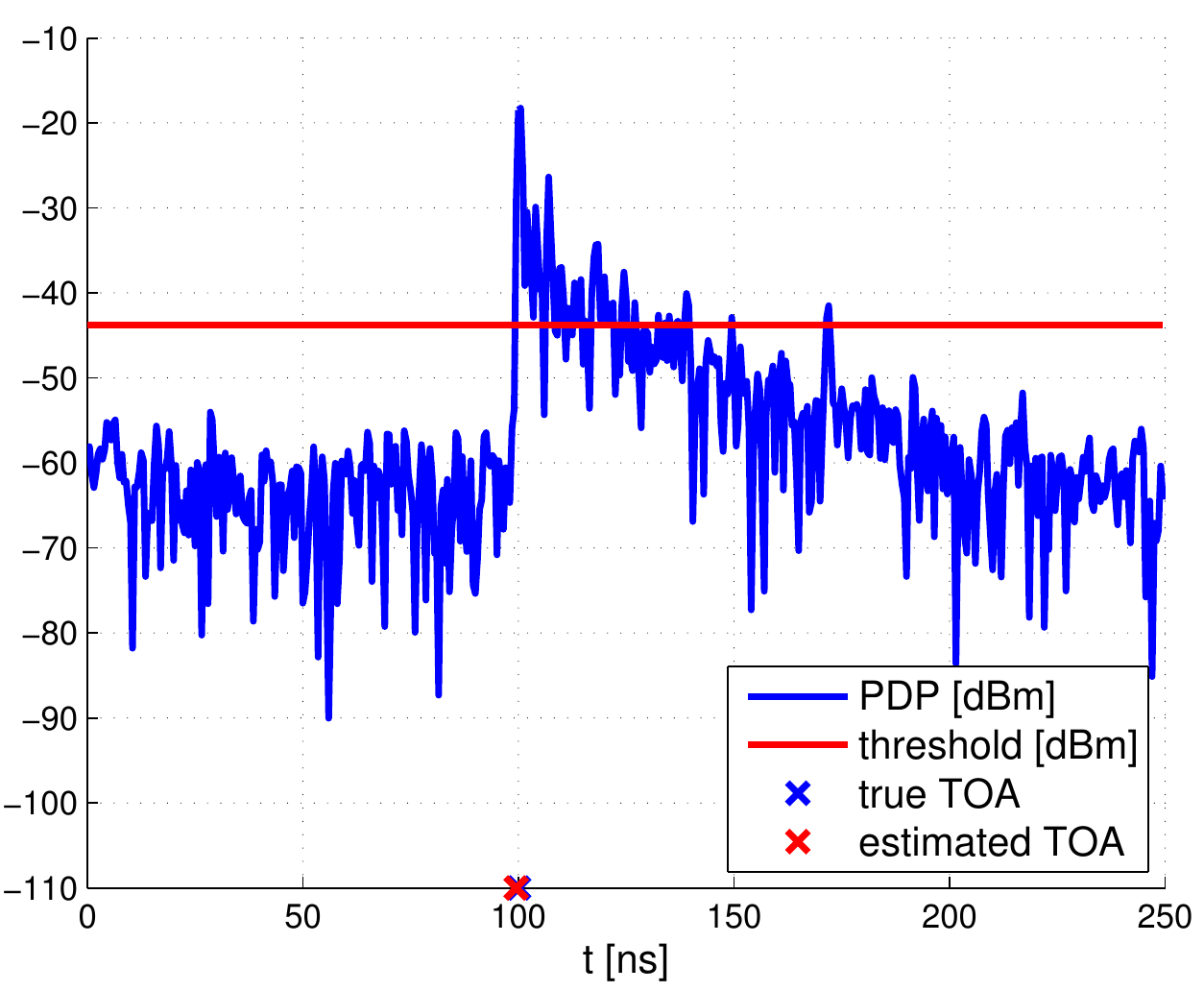}\label{fig:pdpLOS30m}}
\hfill
\subfloat[]{\includegraphics[width=0.51\columnwidth]{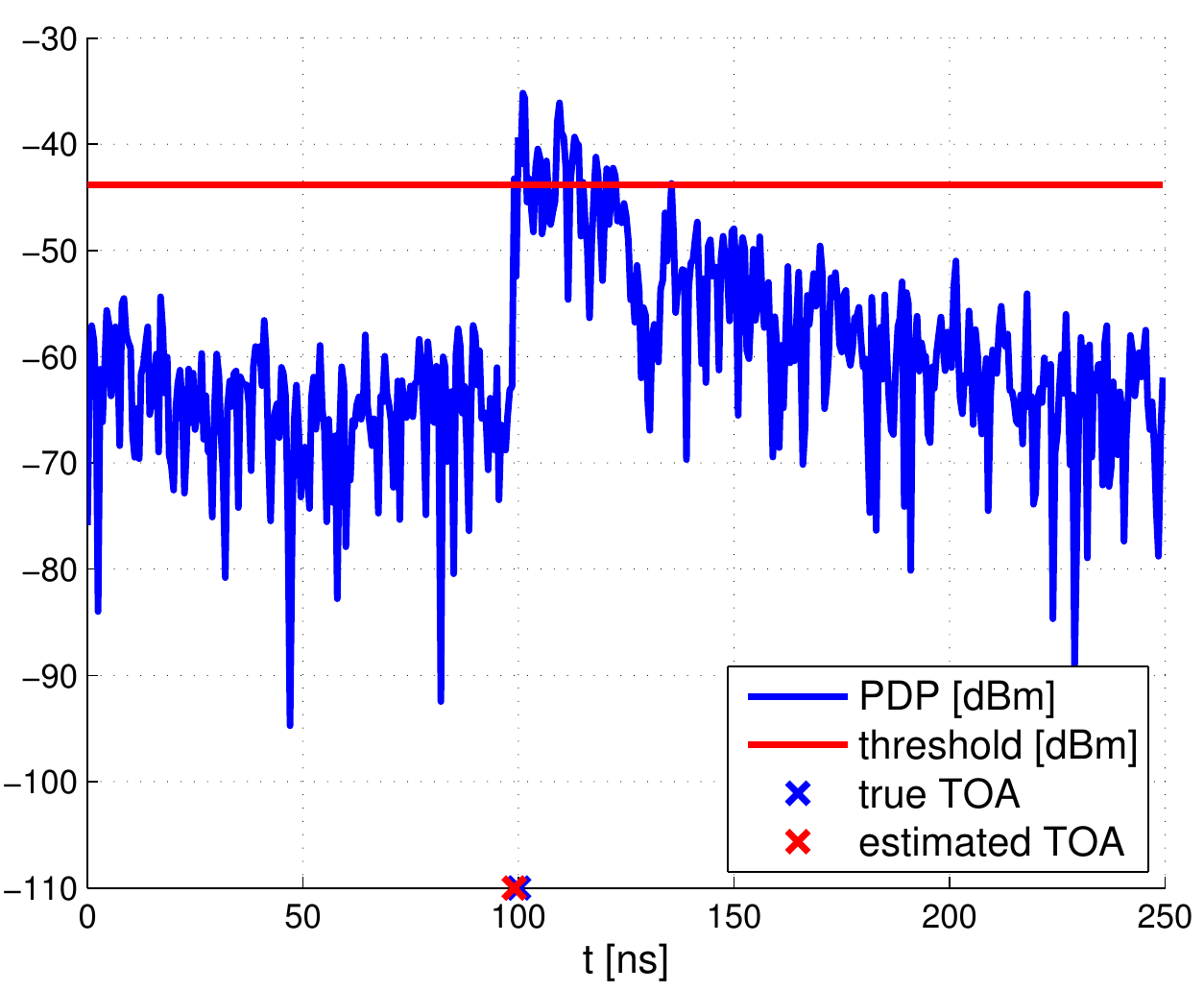}\label{fig:pdpNLOSM30m}}
}
\centerline{
\subfloat[]{\includegraphics[width=0.51\columnwidth]{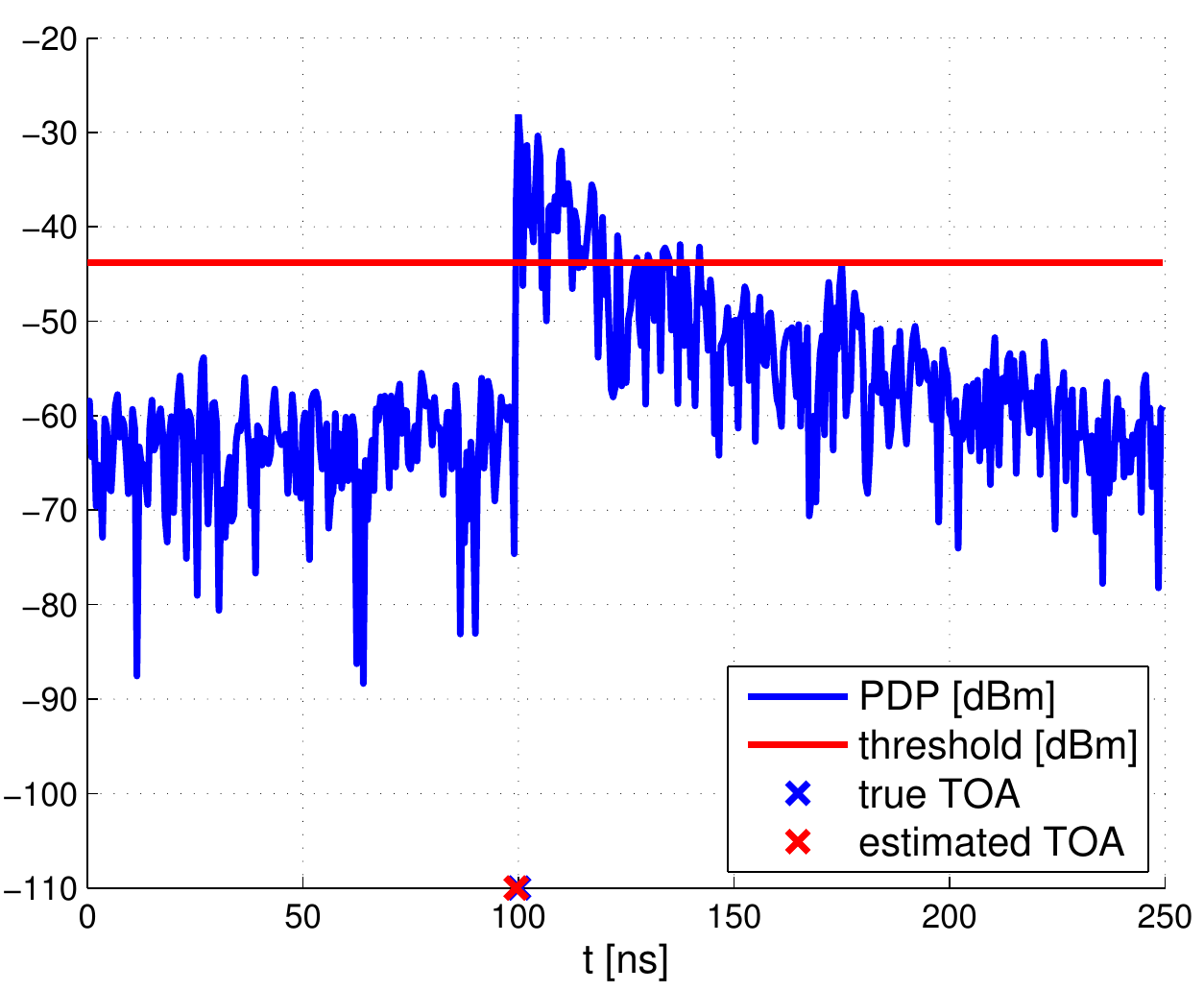}\label{fig:pdpNLOSP30m}}
\hfill
\subfloat[]{\includegraphics[width=0.51\columnwidth]{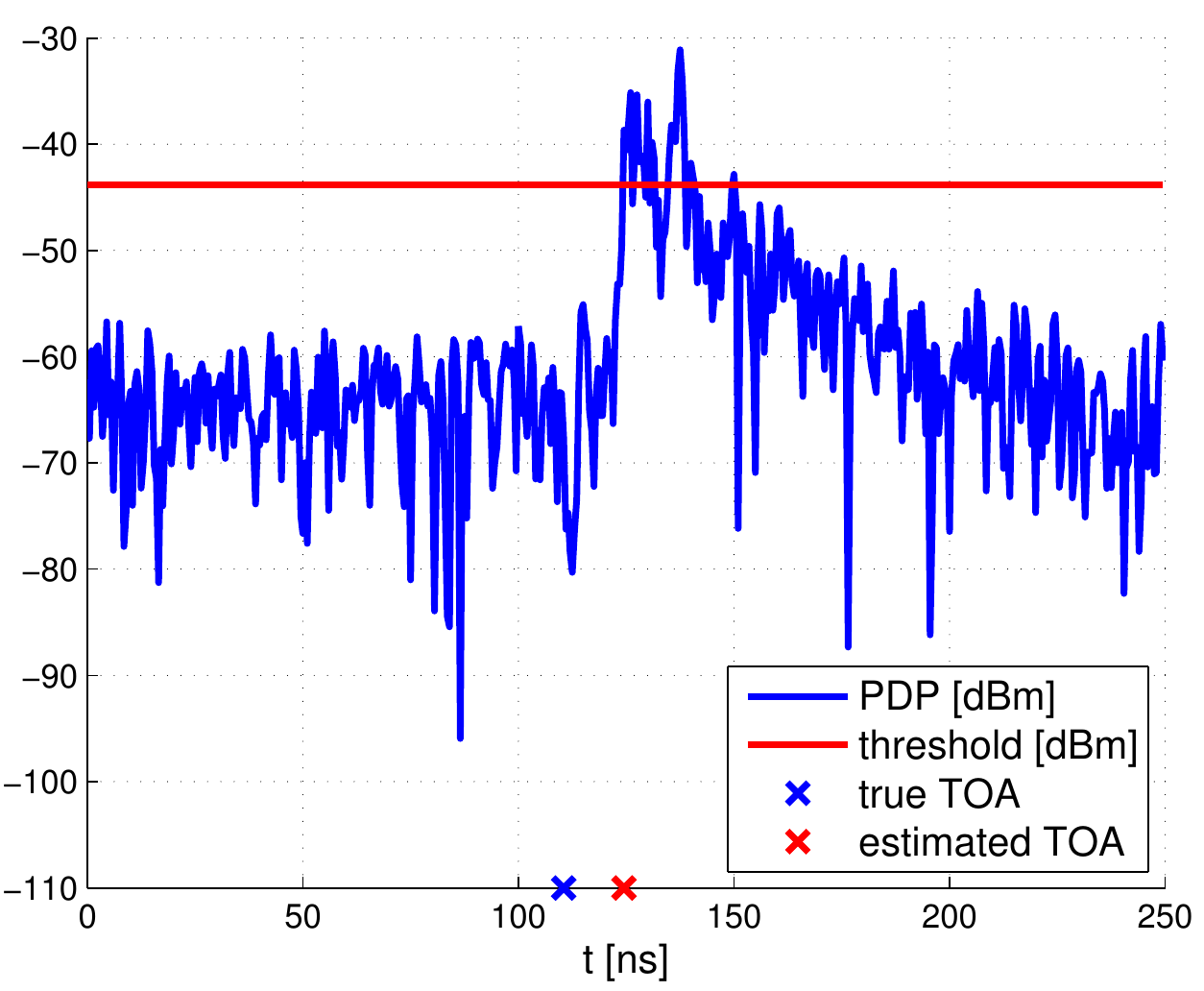}\label{fig:pdpNLOSW33m}}
}
\caption{Illustration of PDPs and threshold-based TOA estimation: (a) LOS, (b) NLOS-M, (c) NLOS-P, and (d) NLOS-W. The power levels are calibrated.}
\label{fig:pdps}
\end{figure}

In Fig. \ref{fig:pdps}, we show an illustration of TOA estimation for all considered scenarios. We can see that TOA estimates are very accurate for all scenarios, except for NLOS-W in which there is a large positive bias. This behavior is expected due to the high resolution of UWB signal, and its capability to penetrate thin obstacles \cite{Gezici2005}.

\section{Measurement analysis and channel modeling for ranging}\label{sec:analysis-model}
In this section, we first analyze the distance estimation using exclusively TOA. Then, we analyze channel propagation parameters in order to determine the most useful ones for NLOS identification and error mitigation. Finally, we propose a statistical model for distance estimation using TOA and the most appropriate channel propagation parameters.

\begin{figure*}[!t]
\centerline{
\subfloat[]{\includegraphics[width=0.33\textwidth]{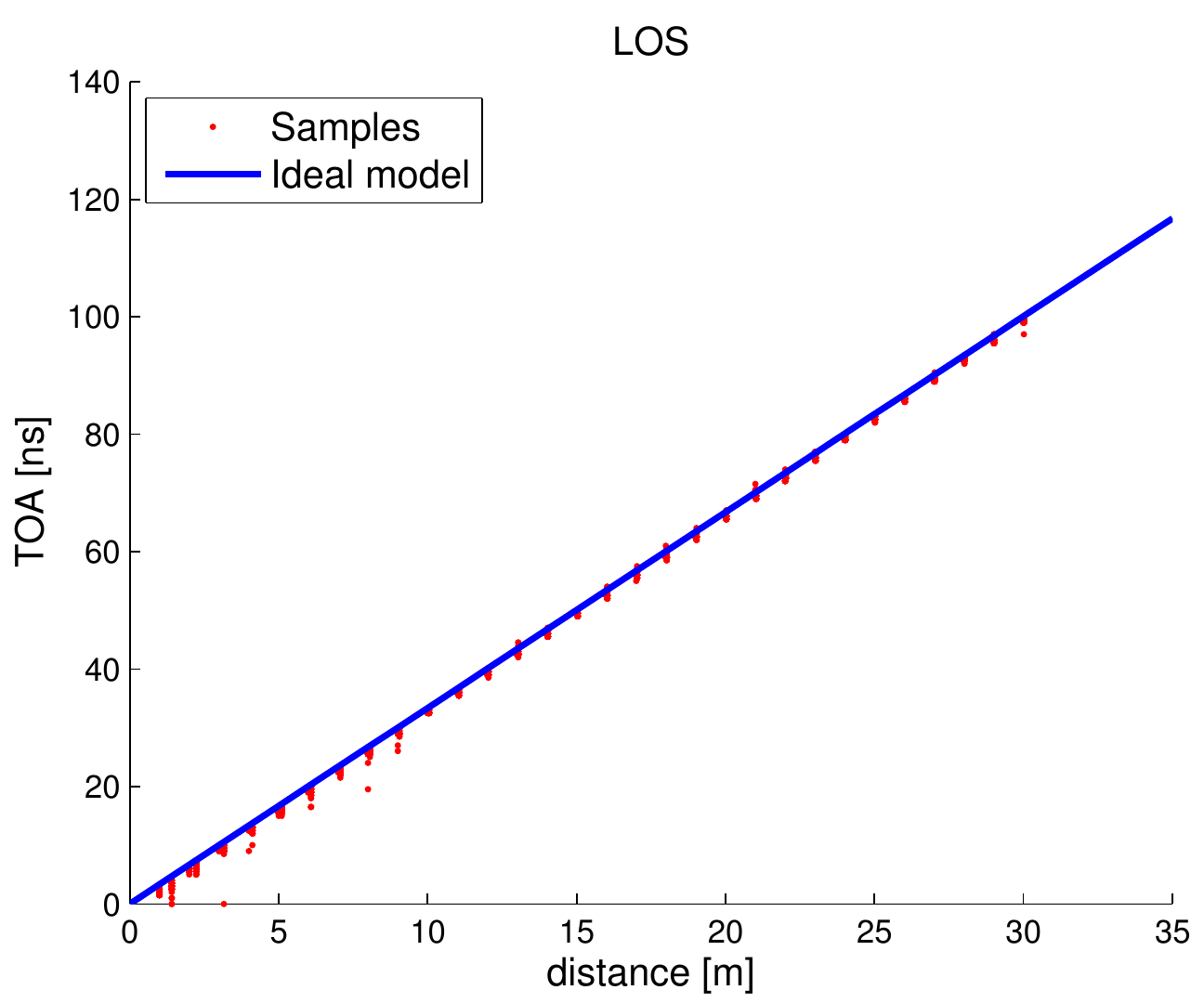}\label{fig:distTOAlos}}
\hfill
\subfloat[]{\includegraphics[width=0.33\textwidth]{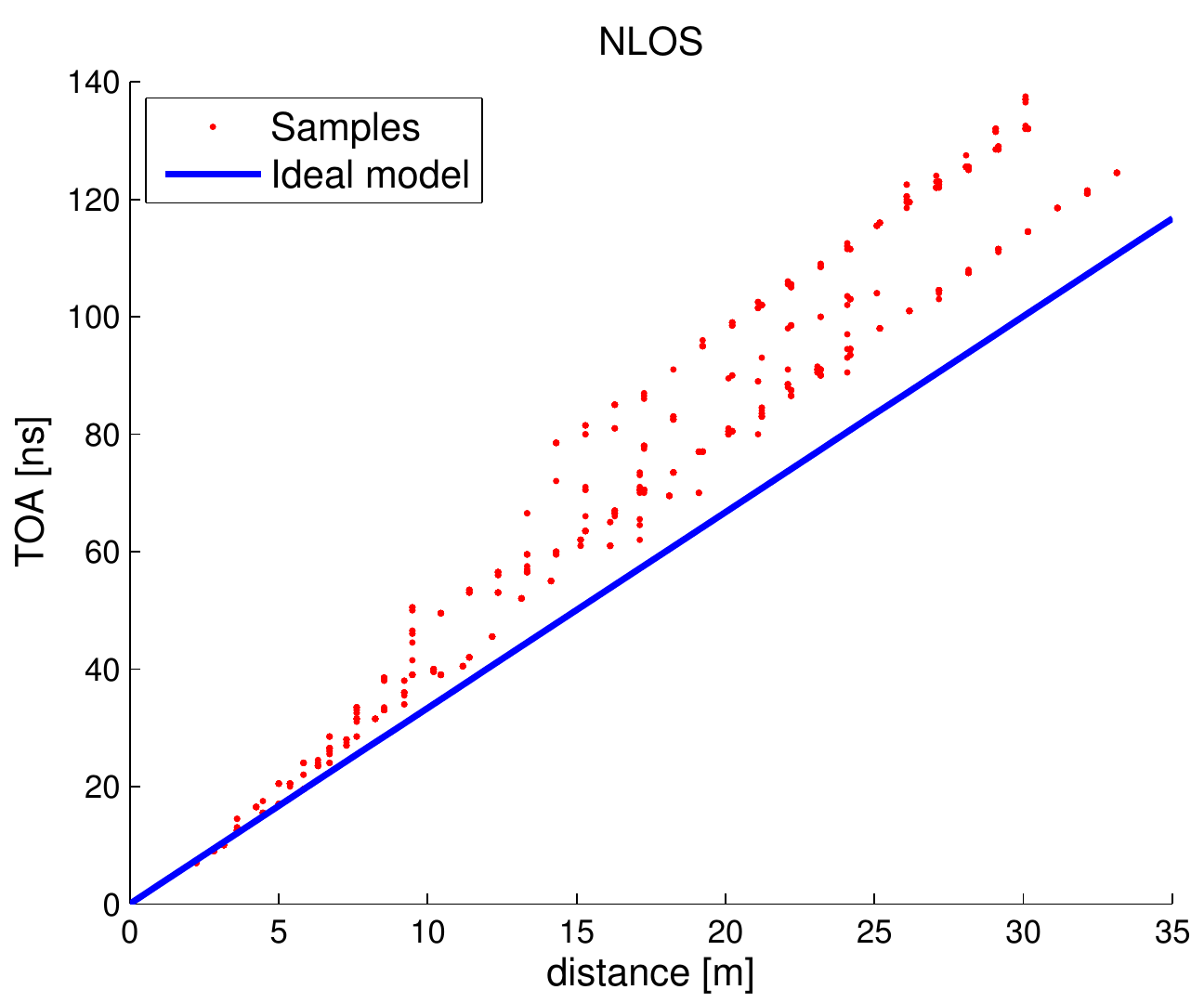}\label{fig:distTOAnlos}}
\hfill
\subfloat[]{\includegraphics[width=0.33\textwidth]{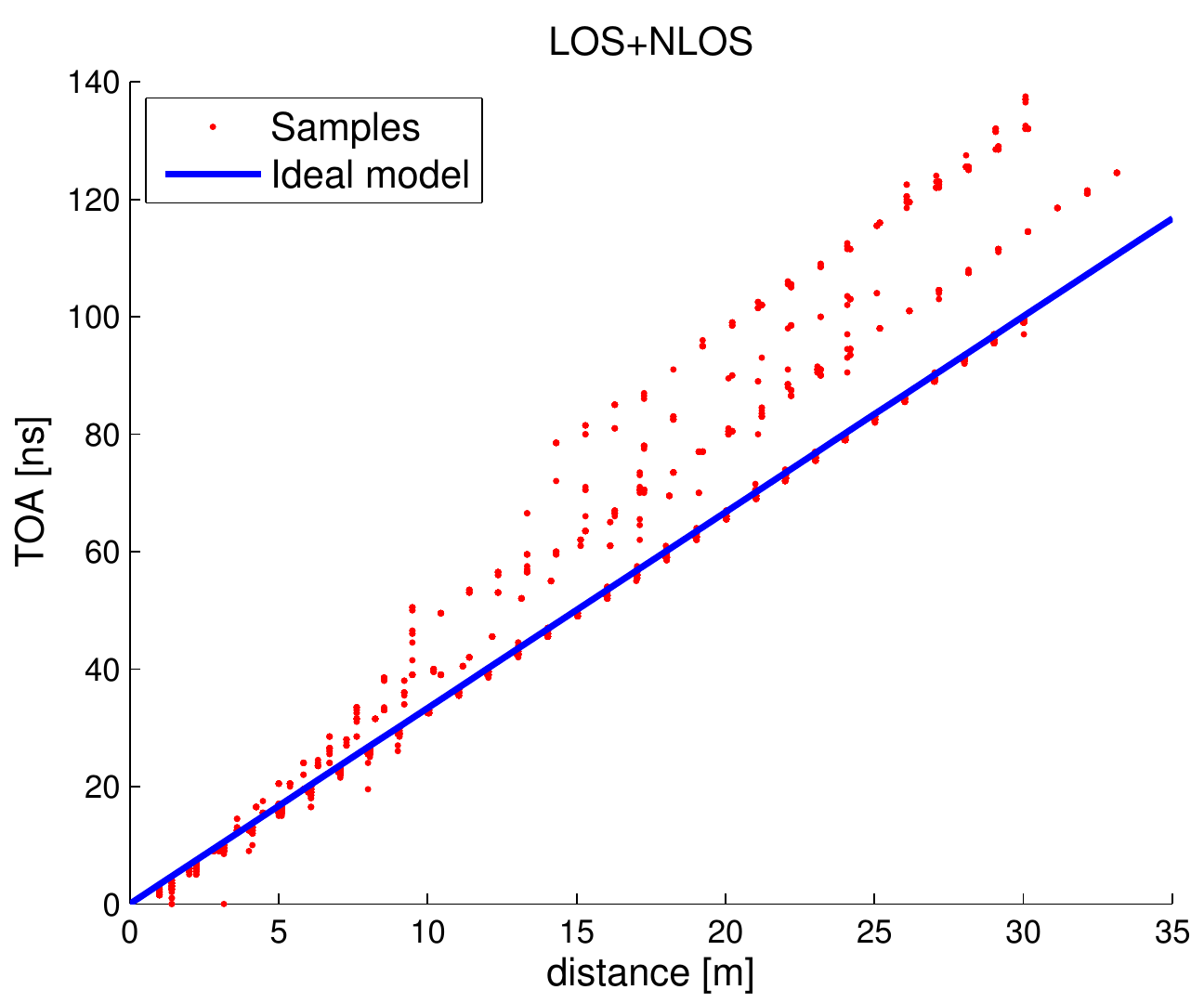}\label{fig:distTOAall}}
}
\centerline{
\subfloat[]{\includegraphics[width=0.33\textwidth]{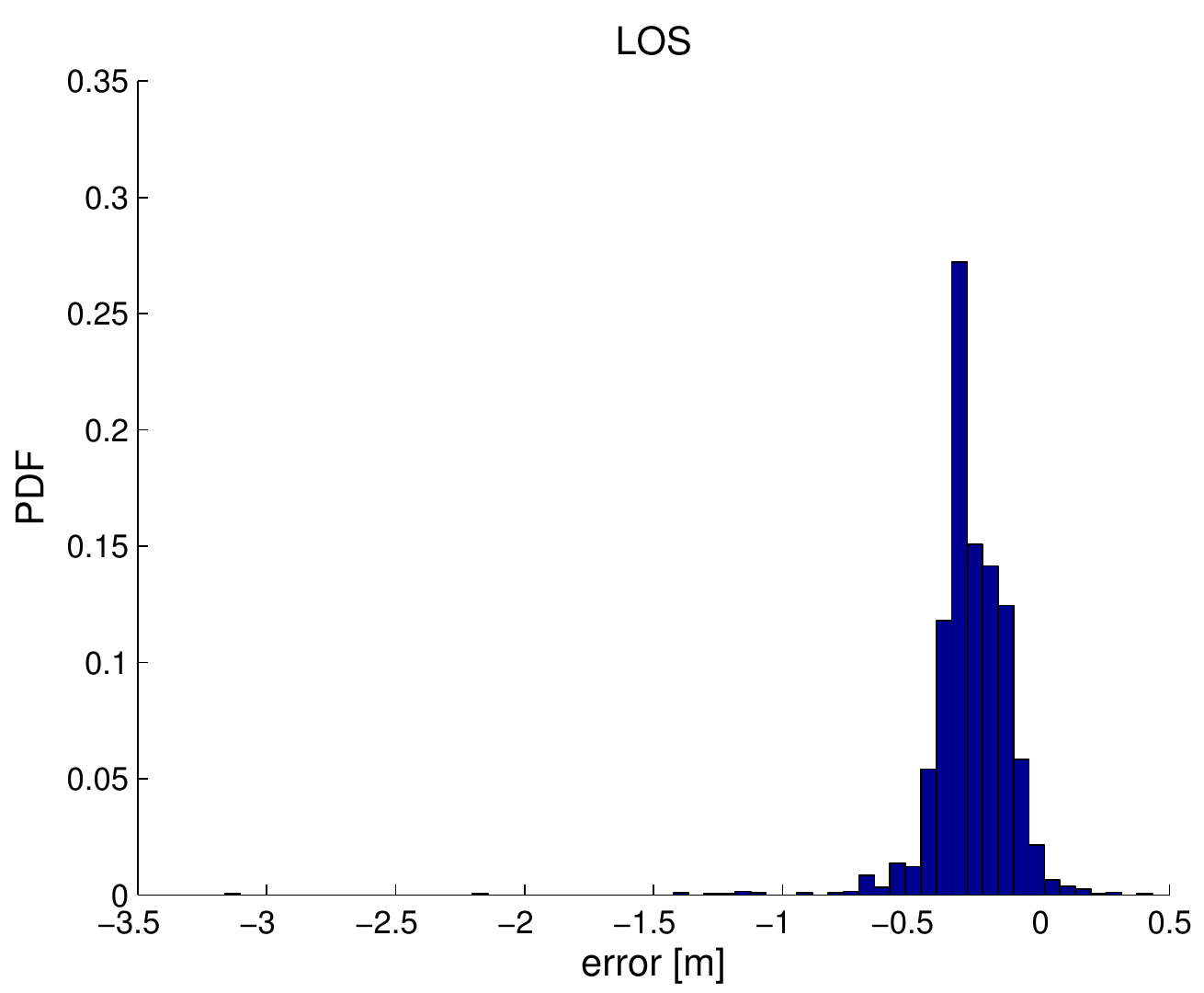}\label{fig:errorTOAlos}}
\hfill
\subfloat[]{\includegraphics[width=0.33\textwidth]{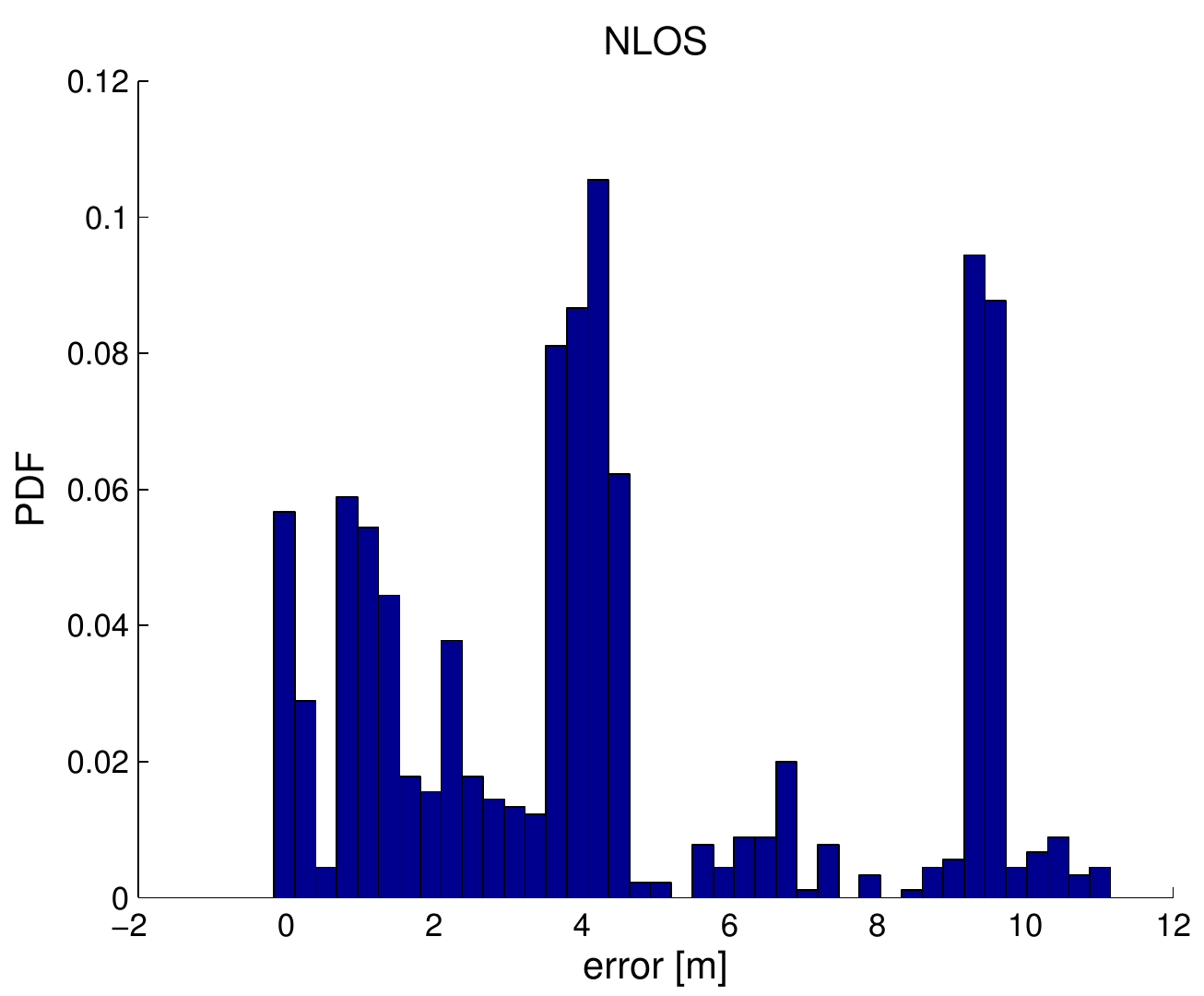}\label{fig:errorTOAnlos}}
\hfill
\subfloat[]{\includegraphics[width=0.33\textwidth]{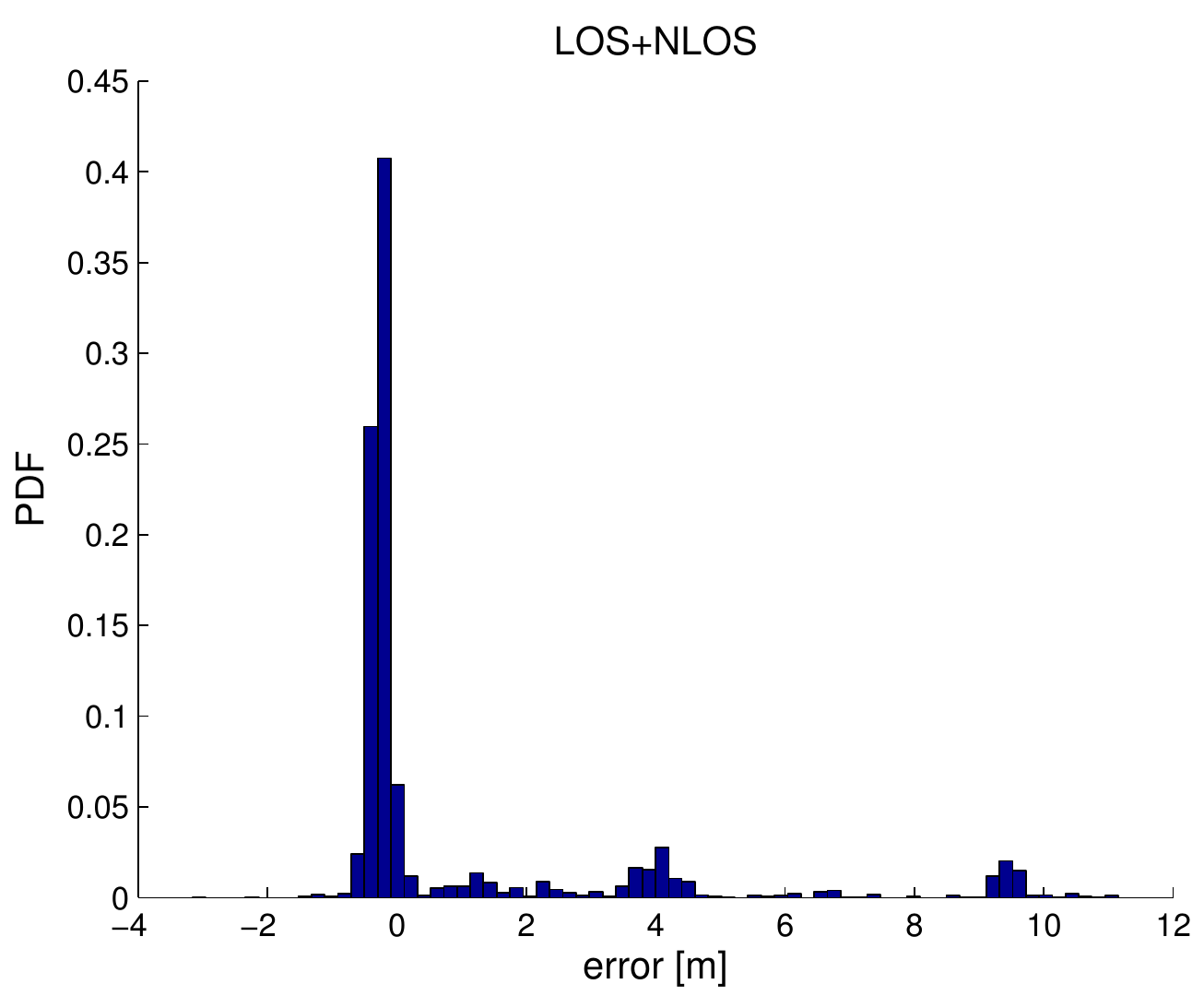}\label{fig:errorTOAall}}
}
\caption{TOA vs.\ distance and error histograms for: (a), (d) LOS (b), (e) NLOS, and (c), (f) LOS+NLOS scenario. The corresponding means and standard deviations of the ranging error are: $\mu_{\nu}=-0.27$ m, $\sigma_{\nu}=0.16$ m (LOS), $\mu_{\nu}=4.38$ m, $\sigma_{\nu}=3.20$ m (NLOS), and $\mu_{\nu}=0.89$ m, $\sigma_{\nu}=2.58$ m (LOS+NLOS).}
\label{fig:toa-dist}
\end{figure*}

\subsection{TOA-based ranging}\label{subsec:toa-ranging}
Since TOA measurements based on a UWB signal can provide very accurate range estimates \cite{Gezici2005} (which will be confirmed in Section \ref{sec:nlos-ident-error}), we will focus on TOA-based ranging. We consider the following model:
\be\label{eq:toa-model}
c\tau_1={d}+\nu
\ee
where $d$ is the true distance between transmitter and receiver, $c=3 \cdot 10^8$ m/s is the speed of light, and $\nu$ represents the measurement noise distributed according to $p_{\nu}(\cdot)$ (with corresponding mean and variance: $\mu_{\nu}$, and $\sigma^2_{\nu}$).

Since our previous analysis showed that LOS, NLOS-P and NLOS-M behave in a similar way (see Fig. \ref{fig:pdps}), we combine them into one sample set. We also consider NLOS-W (in which the direct path is blocked by a wall), and a combination of all sample sets (LOS, NLOS-P, NLOS-M, and NLOS-W). To simplify notation, the combination of LOS and all soft NLOS scenarios (NLOS-P and NLOS-M)  will be referred to as LOS, while NLOS-W will be referred to as NLOS.

The results are shown in Fig. \ref{fig:toa-dist}. As we can see, the TOA estimates provide very accurate distance estimates in the LOS scenario, but there is a large positive error (up to 11 m) in the NLOS scenario. We also note that there are few false alarms caused by a low threshold, but this problem would not appear if there were no losses in the cables (i.e, a higher SNR would allow to use a higher threshold). According to Fig. \ref{fig:toa-dist}(d)-(e), the noise PDF can be approximated with a Gaussian distribution in the case of a LOS scenario, and a Gaussian mixture in the case of an NLOS scenario. Therefore, the model in \eqref{eq:toa-model} is not good enough, and an error mitigation technique is required to enable more accurate ranging.

\subsection{NLOS identification and error mitigation}\label{sec:nlos-ident-error}

We first define a binary variable $H \in \{\rm{LOS}, \rm{NLOS}\}$, and assume that TOA measurement noise is given by:
\be\label{eq:noise-los-nlos}
\nu  = \left\{ \begin{array}{l}
{\nu _L},\,\,{\rm{if}}~{H=\rm{LOS}}\\
{\nu _N},\,{\rm{if}}~{H=\rm{NLOS}}
\end{array} \right. = \left\{ \begin{array}{l}
{\nu _L},\,\,\,\,\,\,\,\,\,\,\,{\rm{if}}~{H=\rm{LOS}}\\
{\nu _L} + b,\,{\rm{if}}~{H=\rm{NLOS}}
\end{array} \right.
\ee
where $\nu_L$ includes all typical sources of the error in the LOS scenario (i.e., thermal noise, finite bandwidth, non-ideal equipment, etc.), and $\nu_N$ in addition includes a positive and random bias $b$ caused by multipath propagation in the NLOS scenario. Since we have available samples of $\nu_N$, we focus on mitigation of the total error.\footnote{To find $b$, we would need to perform a deconvolution, but it is unnecessary since $b>>\nu_L$ according to Fig. \ref{fig:toa-dist}.} 

Our goal is to identify the channel state (estimate $H$), and to remove (or at least, reduce) the NLOS error. Thus, we need to choose an appropriate NLOS identification and error mitigation technique. Since we would like to use one single impulse response, to keep the  complexity reasonable, and to avoid using the geometry of the tunnel, we will apply the algorithm based on channel propagation parameters (for alternatives, see Section \ref{sec:nlos-overview}). Although state-of-the-art (e.g., see \cite{Venkatesh2007a,Zhang2013,Denis2003}) already provides  parameters that are useful for these problems, they may not be the best for tunnel environment. Therefore, in order to determine which parameters are appropriate, we use the following metrics:
\bi
\item \textit{Overlap metric} \cite{Venkatesh2007a} for parameter ${\alpha}$ ($0<\xi_{\alpha}<\infty$) :
\be\label{eq:overlap}
\xi_{\alpha}=\frac{\sqrt{\sigma_{\alpha_L}\sigma_{\alpha_N}}}{\left|\mu_{\alpha_N}-\mu_{\alpha_L}\right|}
\ee
 where ${\alpha}$ can be any of the channel propagation parameters defined in eqs. \eqref{eq:toa}-\eqref{eq:kurt}; $\mu_{\alpha_L}$, $\sigma_{\alpha_L}$, and $\mu_{\alpha_N}$, $\sigma_{\alpha_N}$ represent the means and the standard deviations of PDFs $p(\alpha|H={\rm{LOS}})$ and $p(\alpha|H={\rm{NLOS}})$, respectively. They are obtained from an appropriate sample set. A smaller value of $\xi_{\alpha}$ implies that there is less overlap between the LOS and NLOS distributions, so we can more easily distinguish between the LOS and NLOS states.
 \item Correlation coefficient between the parameter ${\alpha}$ and the true distance $d$ ($-1<{\rho_{\alpha ,d}}<1$):
 \be\label{eq:corr-dist}
{\rho _{\alpha ,d}} = \frac{{{\rm{Cov}}(\alpha ,d)}}{{{\sigma _\alpha }{\sigma _d}}}
 \ee
 where ${{\rm{Cov}}(\alpha ,d)}$, $\sigma _\alpha$, and $\sigma _d$ are computed using all available LOS and NLOS samples. For NLOS identification, a smaller $\left|{\rho _{\alpha ,d}}\right|$ is preferable since the true distance is unknown.
\item Correlation coefficient between the parameter ${\alpha}$ and the NLOS error $\nu_N$ ($-1<{\rho_{\alpha ,\nu_N}}<1$):
\be\label{eq:corr-error}
{\rho _{\alpha ,\nu_N}} = \frac{{{\rm{Cov}}_N(\alpha ,\nu_N)}}{{{\sigma _{\alpha_N} }{\sigma _{\nu_N}}}}
\ee
where ${{\rm{Cov}}_N(\alpha ,\nu_N)}$, $\sigma _{\alpha_N}$, and $\sigma _{\nu_N}$ are computed using only the NLOS sample set. A larger $\left|{\rho _{\alpha ,\nu_N}}\right|$ implies that the error can be more easily determined from the parameter $\alpha$.
\item Correlation coefficient between the two parameters ${\alpha_1}$ and ${\alpha_2}$ ($-1<{\rho_{\alpha_1 ,\alpha_2}}<1$):
\be\label{eq:corr-param}
{\rho_{\alpha_1 ,\alpha_2}} = \frac{{{\rm{Cov}}(\alpha_1,\alpha_2)}}{{{\sigma _{\alpha_1} }{\sigma _{\alpha_2}}}}
\ee
\ei
 where ${{\rm{Cov}}(\alpha_1,\alpha_2)}$, $\sigma _{\alpha_1}$, and $\sigma _{\alpha_2}$ are computed using all available LOS and NLOS samples. A large value of $\left|{\rho_{\alpha_1 ,\alpha_2}}\right|$ means that one of the parameters can be discarded.
 
Note that we will not take into account very small differences between obtained values since we have a limited set of data (2700 LOS and 900 NLOS samples).

\begin{table*}[!tb]
\caption{(a) Estimated $\xi_{\alpha}$, ${\rho _{\alpha ,\nu_N}}$ and ${\rho _{\alpha ,d}}$ for all considered parameters, and (b) estimated ${\rho_{\alpha_1 ,\alpha_2}}$ between all pairs of the parameters. High levels of absolute correlation ($>0.7$) and overlap ($>2$) are marked with red, while low levels of absolute correlation ($<0.3$) and overlap ($<1$) are marked with blue color.}
\label{table:metrics}
\centering
\subfloat[]{
\begin{tabular}{c||c|c|c}
 ${\alpha}$ & $\xi_{\alpha}$ &  ${\rho _{\alpha ,d}}$ & ${\rho _{\alpha ,\nu_N}}$ \\
 \hline\hline
${\tau _1}$ & 1.65 & \red{0.97} 	& \red{0.85} \\\hline
$P_{RSS}$ 	& 1.10 & \red{-0.81} 	& \red{-0.73} \\\hline
$P_{MAX}$ 	& 1.07 &  \red{-0.72} & -0.65 \\\hline
$\bar\tau$ 	& 1.07 &  \red{0.93} 	& \red{0.81} \\\hline
${\tau_{MAX}}$ 	& \red{2.12} &  -0.68 & \red{-0.84} \\\hline
${\tau _{RMS}}$ & \red{3.08} &  -0.55 & \red{-0.81} \\\hline
${\tau _{RT}}$ 	& \blue{0.58} &  \blue{-0.14} & -0.42\\\hline
$\kappa$ 		& \red{4.80} &  -0.48 & -0.44 \\\hline
\end{tabular}}
\hfill
\subfloat[]{
\begin{tabular}{c||c|c|c|c|c|c|c|c}
& ${\tau _1}$ & $P_{RSS}$ & $P_{MAX}$ & $\bar\tau$ & ${\tau_{MAX}}$ & ${\tau _{RMS}}$ & ${\tau _{RT}}$ & $\kappa$ \\
 \hline\hline   
${\tau _1}$ 	& \red{1.00} & \red{-0.83} & \red{-0.76} & \red{0.98} & -0.69 & -0.55 & \blue{-0.06} & -0.49 \\\hline
$P_{RSS}$ 		& -    & \red{1.00}  & \red{0.97}  &\red{-0.88} &  \red{0.79} &  \blue{0.22} & \blue{-0.18} &  0.65 \\\hline
$P_{MAX}$ 		& -    & -     & \red{1.00}  &\red{-0.83} &  \red{0.72} &  \blue{0.12} & \blue{-0.25} &  \red{0.75} \\\hline
$\bar\tau$ 		& -    & -     & -     & \red{1.00} & -0.69 & -0.43 &  \blue{0.11} & -0.52 \\\hline
${\tau_{MAX}}$ 	& -    & -     & -     & -    &  \red{1.00} &  0.31 &  \blue{0.03} &  0.45 \\\hline
${\tau _{RMS}}$ & -    & -     & -     & -    & -     & \red{1.00}  &  0.41 & \blue{0.11}  \\\hline
${\tau _{RT}}$ 	& -    & -     & -     & -    & -     & -     & \red{1.00}  & \blue{-0.11} \\\hline
$\kappa$ 		& -    & -     & -     & -    & -     & -     & -     & \red{1.00}  \\\hline
\end{tabular}}
\end{table*}

\begin{figure*}[!tb]
\centerline{
\subfloat[]{\includegraphics[width=0.31\textwidth]{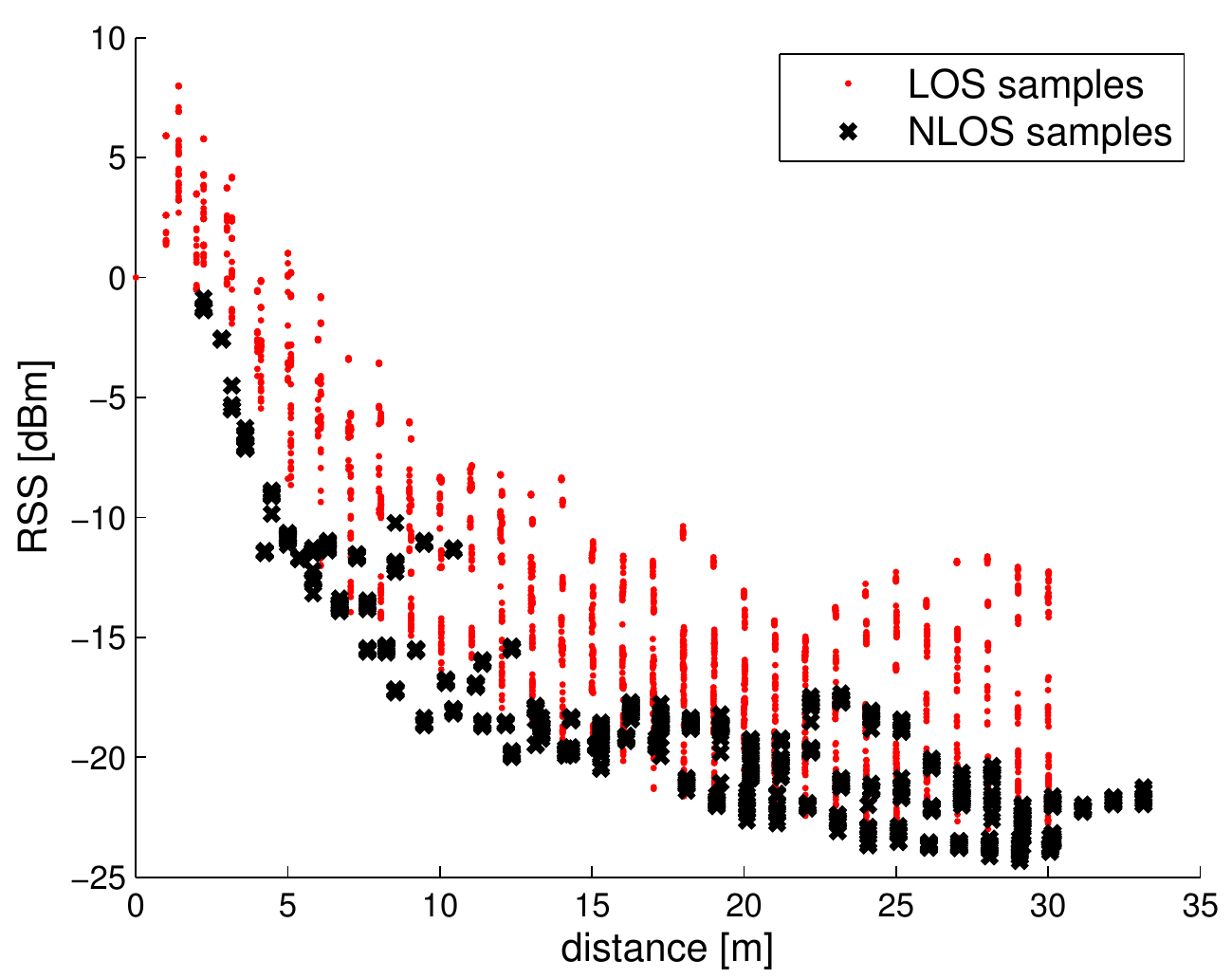}\label{fig:rssDistAll}}
\hfill
\subfloat[]{\includegraphics[width=0.31\textwidth]{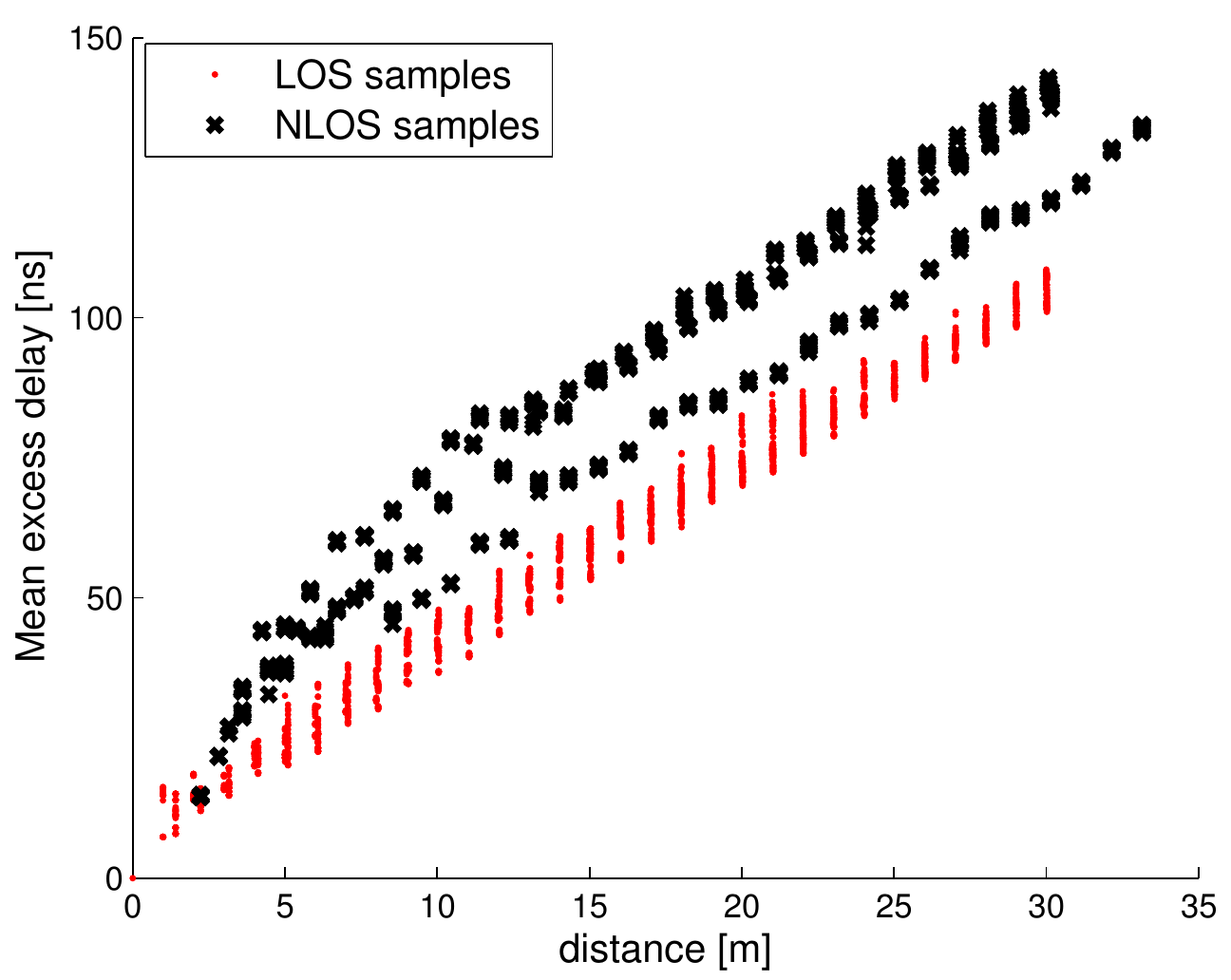}\label{fig:mdsDistAll}}
\hfill
\subfloat[]{\includegraphics[width=0.31\textwidth]{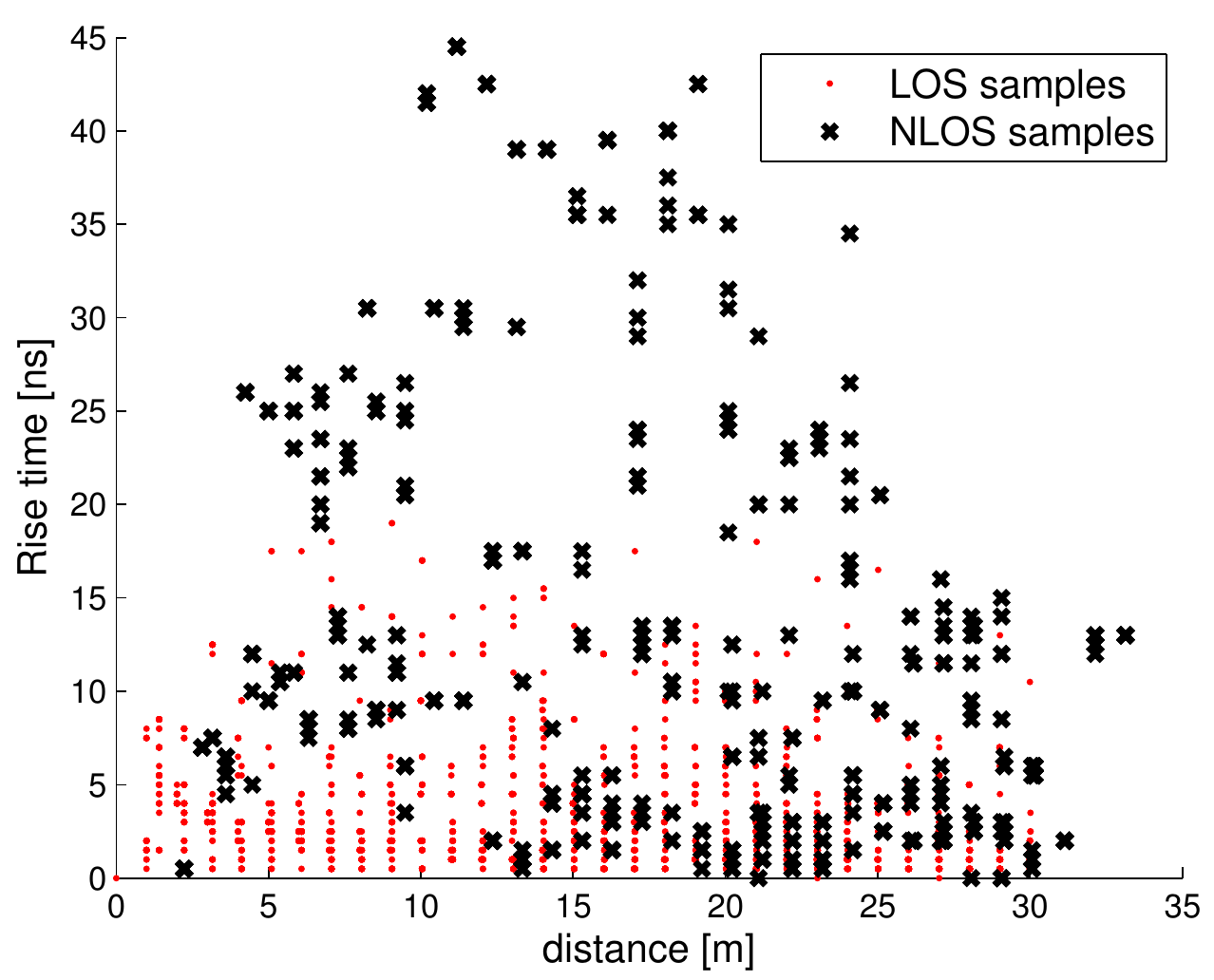}\label{fig:rtDistAll}}
}
\centerline{
\subfloat[]{\includegraphics[width=0.31\textwidth]{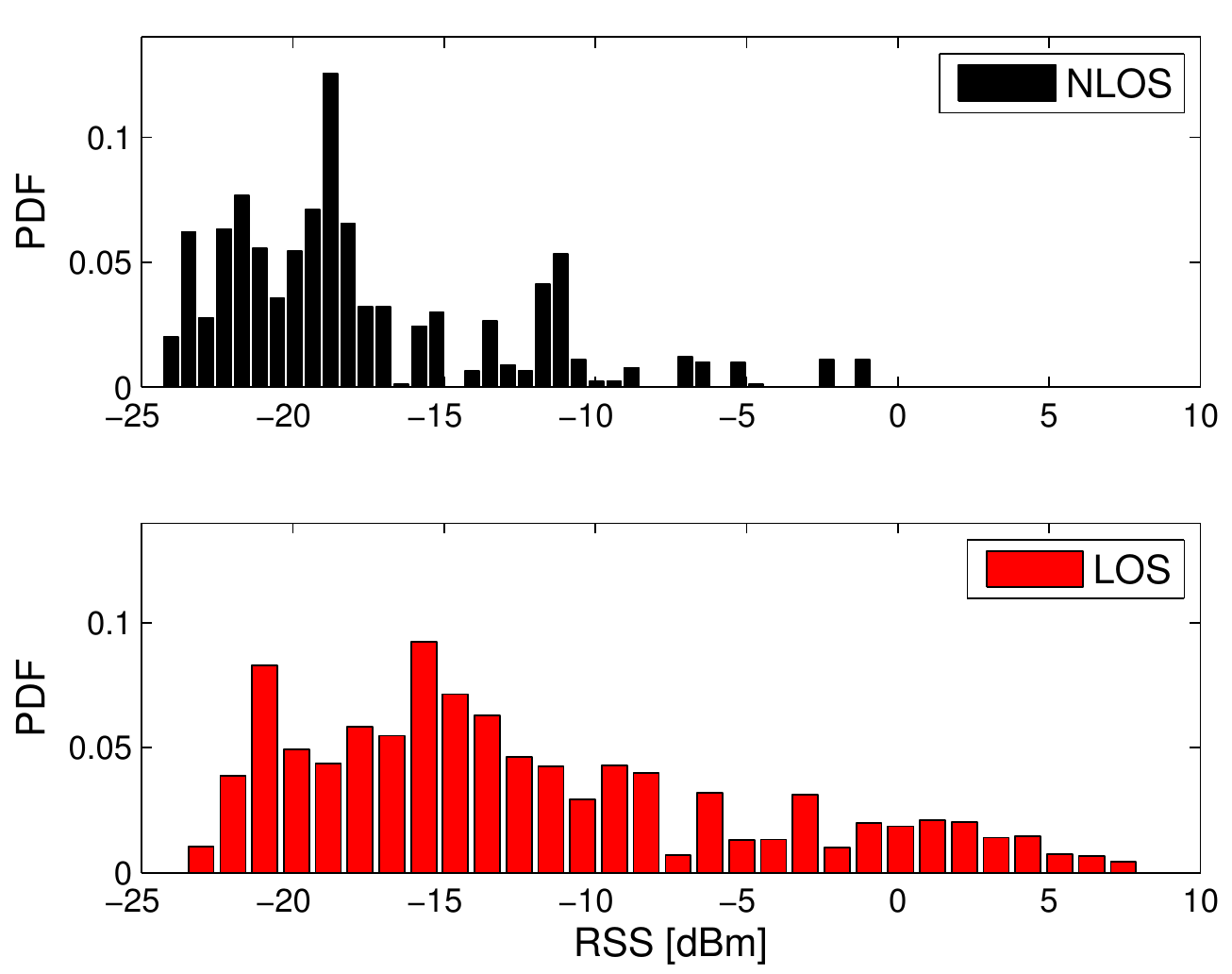}\label{fig:rssHistAll}}
\hfill
\subfloat[]{\includegraphics[width=0.31\textwidth]{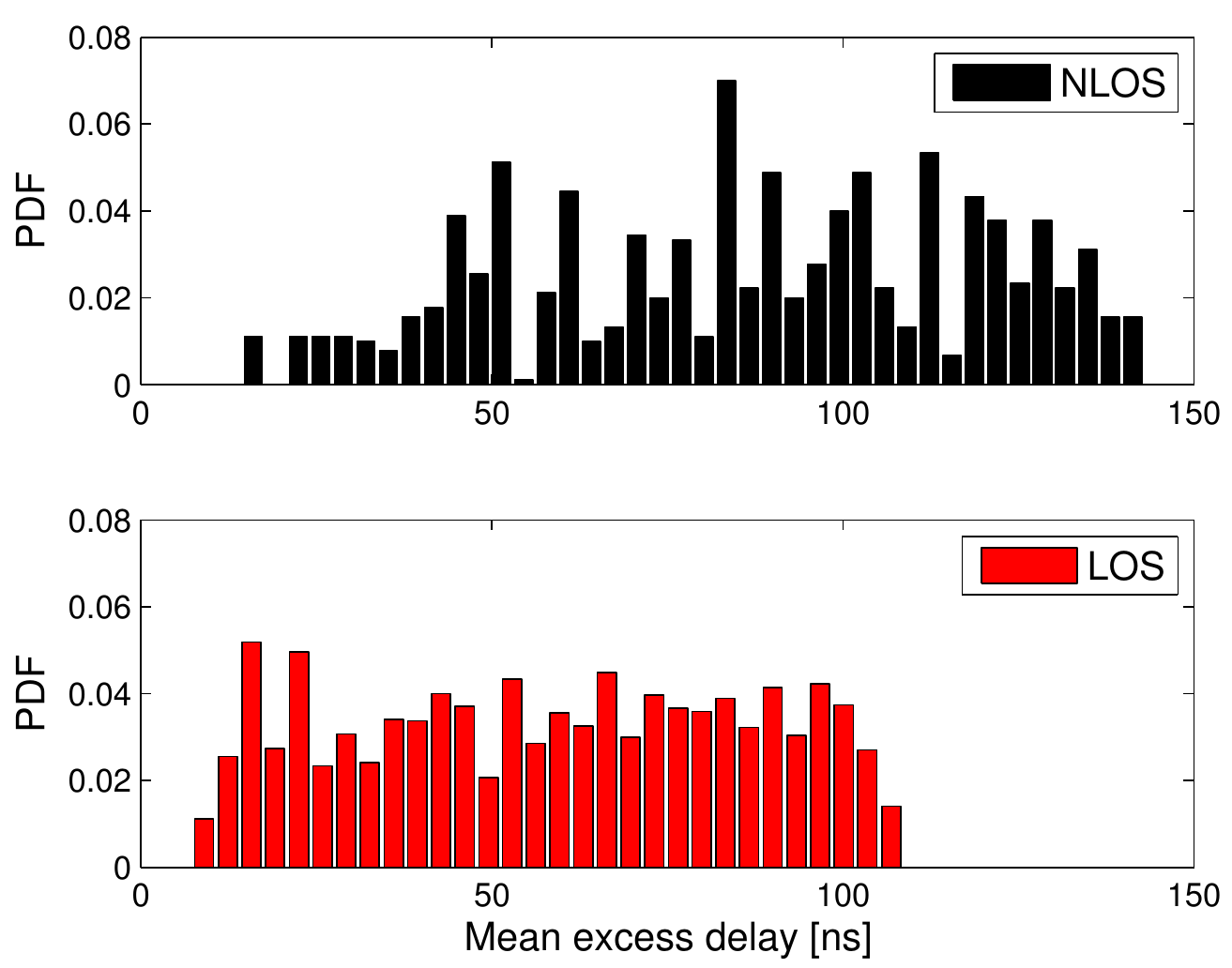}\label{fig:mdsHistAll}}
\hfill
\subfloat[]{\includegraphics[width=0.31\textwidth]{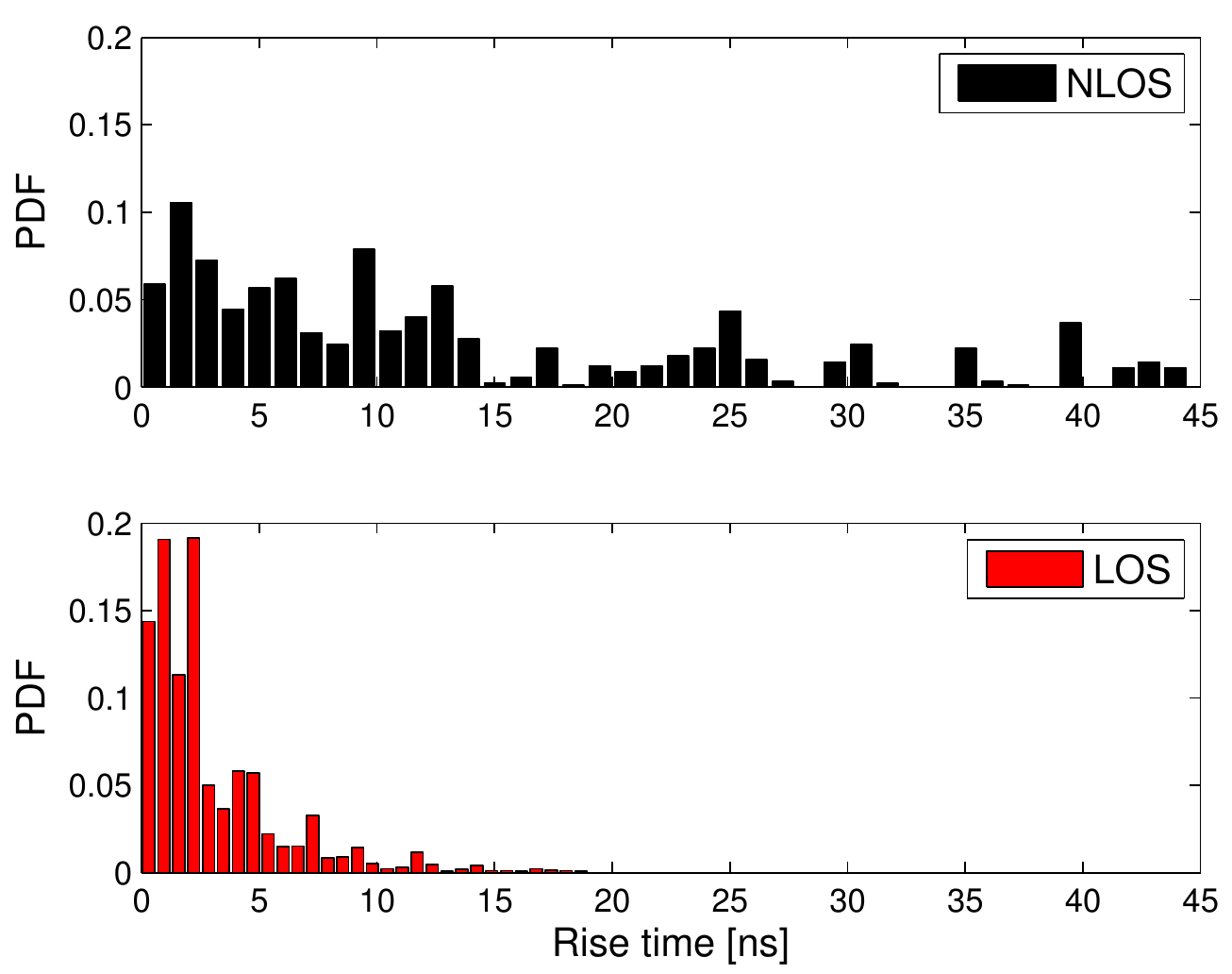}\label{fig:rtHistAll}}
}
\caption{Channel propagation parameters (a) $P_{RSS}$, (b) $\bar\tau$, and (c) ${\tau _{RT}}$, as a function of true distance. (d)-(f) LOS and NLOS histograms corresponding to samples from (a)-(c).}
\label{fig:nlos-ident}
\end{figure*}

\begin{figure*}[!thb]
\centerline{
\subfloat[]{\includegraphics[width=0.30\textwidth]{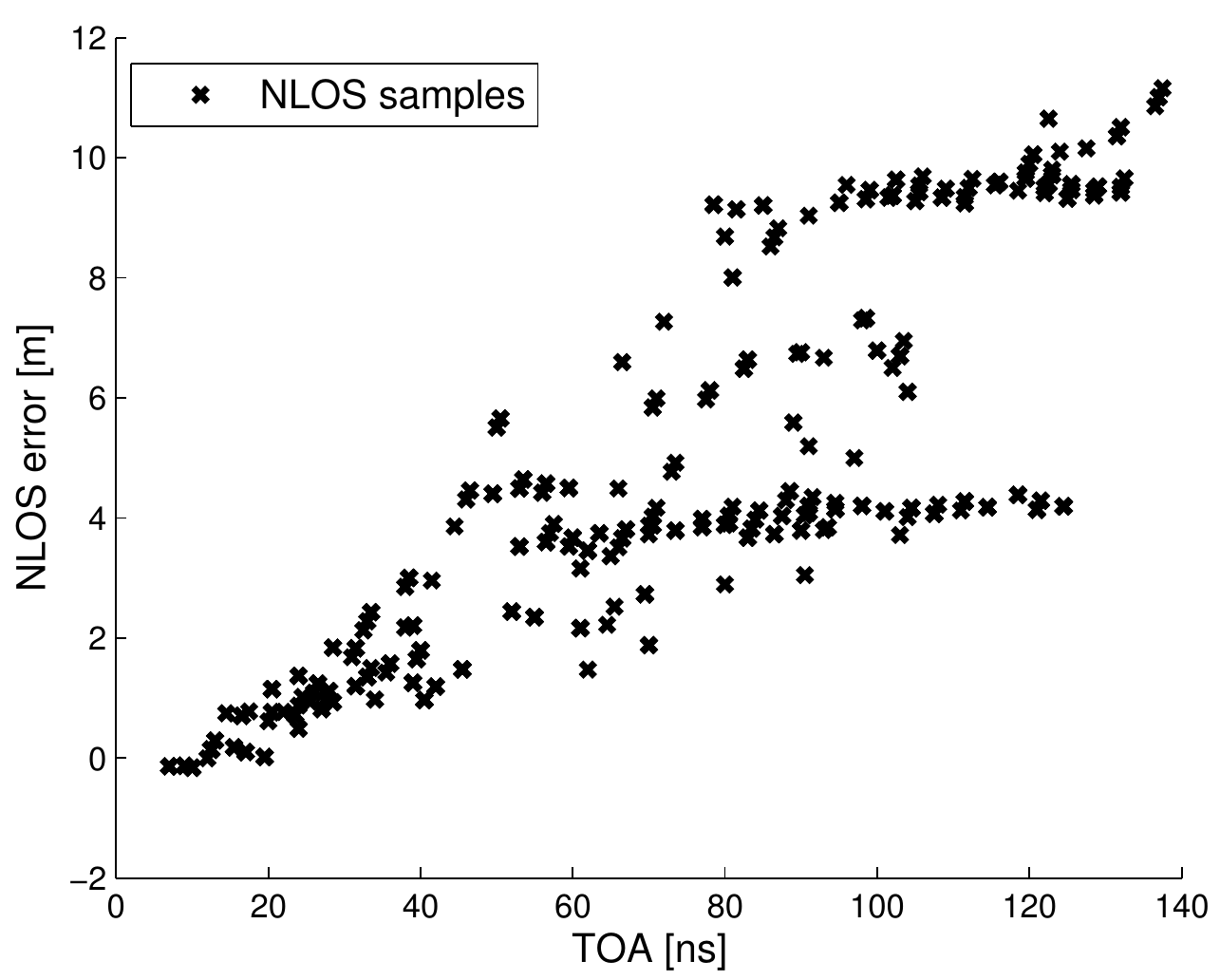}\label{fig:errorTOAnlos}}
\hfill
\subfloat[]{\includegraphics[width=0.30\textwidth]{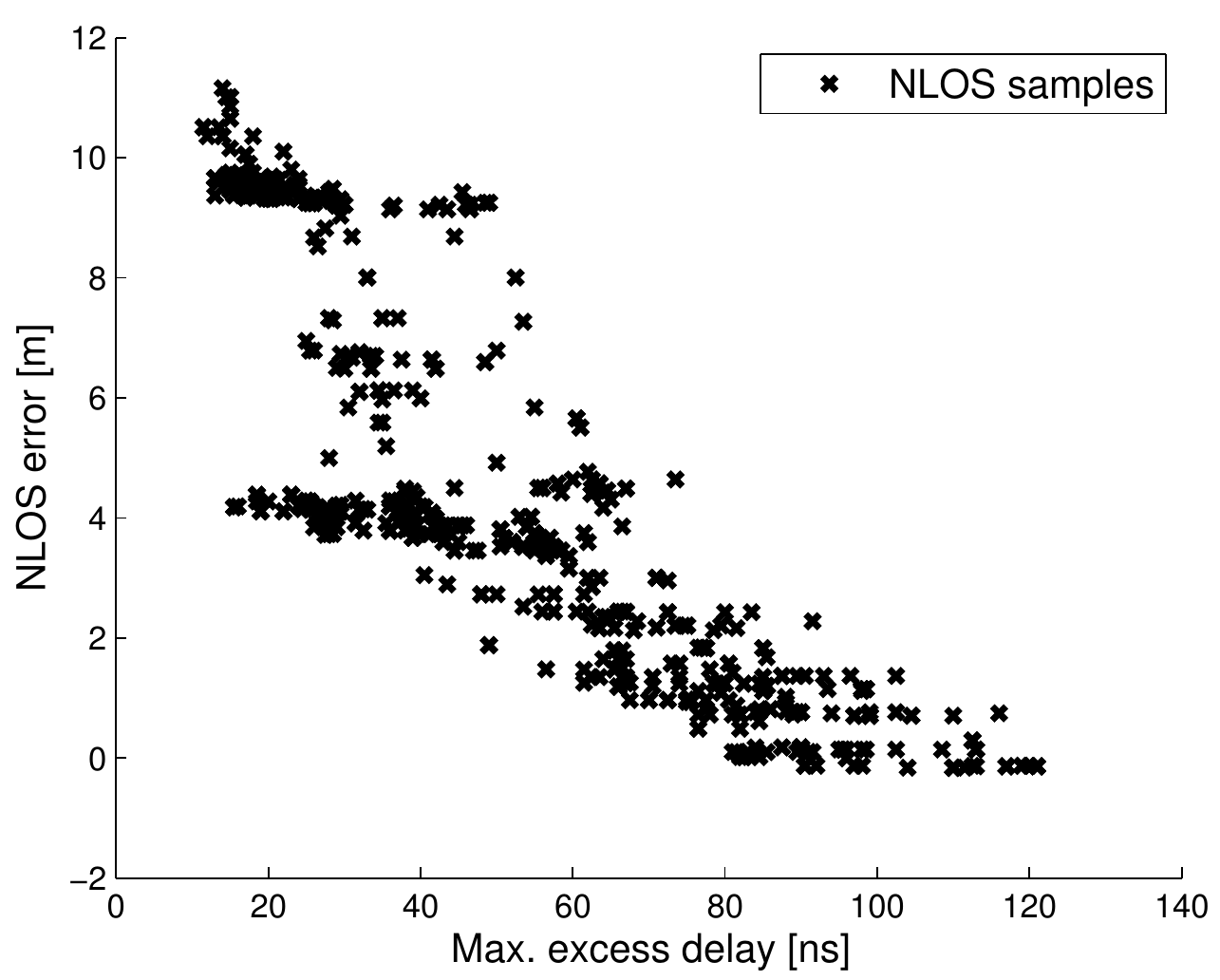}\label{fig:errorMDSnlos}}
\hfill
\subfloat[]{\includegraphics[width=0.30\textwidth]{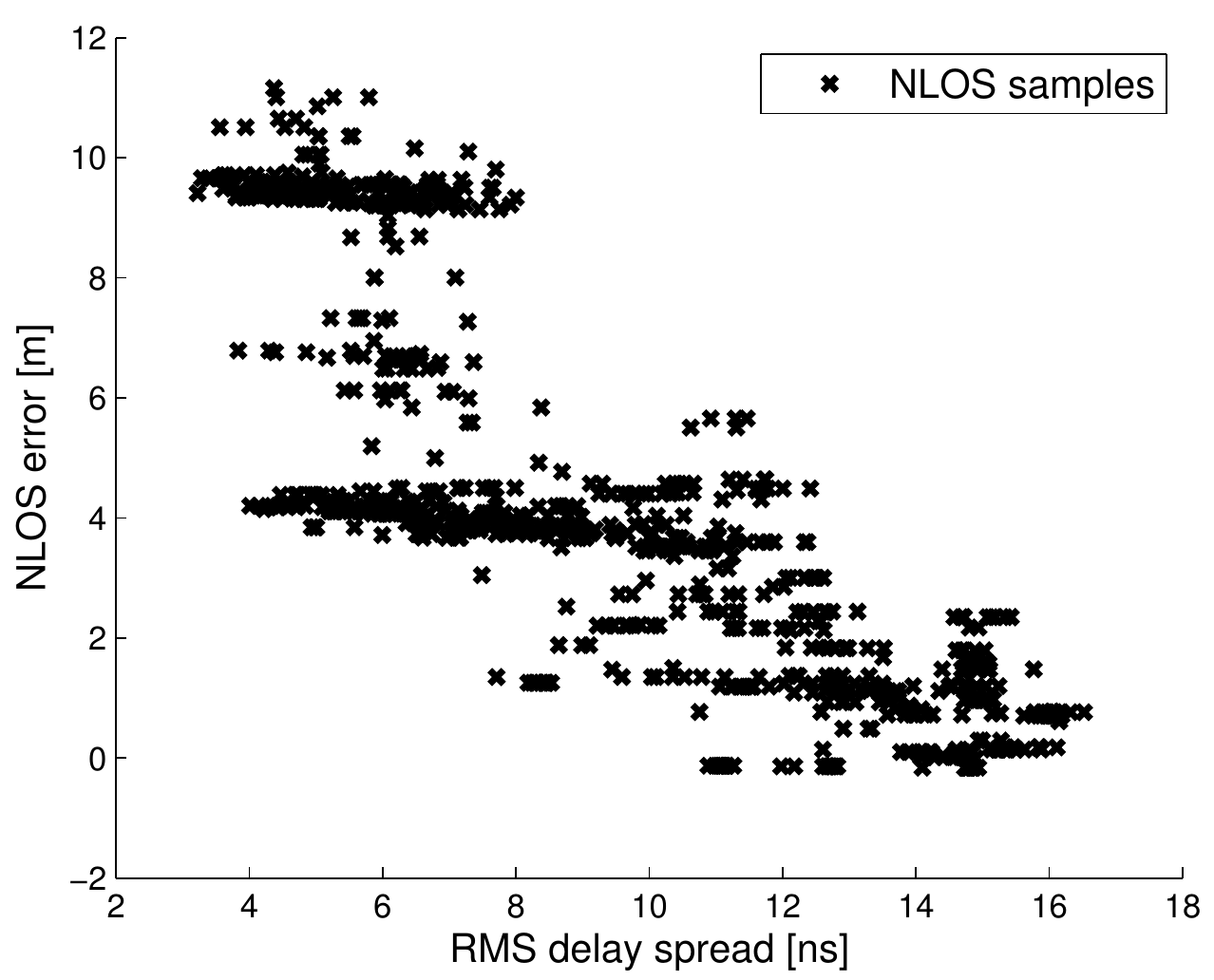}\label{fig:errorRDSnlos}}
}
\caption{NLOS error as function of: (a) ${\tau _1}$, (b) ${\tau_{MAX}}$, and (c) ${\tau_{RMS}}$.}
\label{fig:error-mitig}
\end{figure*}

Estimated results are shown in Tables \ref{table:metrics}. We observe the following:
\bi
\item ${\tau _1}$ is strongly correlated with distance, so we can confirm that it is the best parameter for distance estimation (see also Section \ref{subsec:toa-ranging}). Good options are also $P_{RSS}$, $P_{MAX}$, and $\bar\tau$, but they are strongly correlated with ${\tau _1}$, and between each other.
\item ${\tau _{RT}}$ provides the lowest overlap, so it is the best parameter for NLOS identification. It is also uncorrelated with the true distance, which means that a simple identification algorithm can be used. One of the parameters, $P_{RSS}$, $P_{MAX}$, or $\bar\tau$, could be also used since they are uncorrelated with ${\tau _{RT}}$, but they are strongly correlated with true distance. We also note that ${\tau _{RMS}}$ and $\kappa$ are the worst parameters for this problem in contrast to some other results in literature (e.g., \cite{Venkatesh2007a,Zhang2013}).
\item Due to the strong correlation with the NLOS error, the best parameters for NLOS error mitigation are: ${\tau _1}$, $P_{RSS}$, $\bar\tau$, ${\tau_{MAX}}$, and ${\tau _{RMS}}$. However, $P_{RSS}$ and $\bar\tau$ can be discarded since they are strongly correlated with ${\tau _1}$. Out of the remaining parameters, ${\tau _{RMS}}$ and ${\tau_{MAX}}$ are the most appropriate due to their lower correlation with the true distance.
\ei

We illustrate, in Fig. \ref{fig:nlos-ident} and Fig. \ref{fig:error-mitig},  the results for the most competitive parameters. 

As we can observe in Fig. \ref{fig:nlos-ident}, $\bar\tau$ can be used to (almost) perfectly detect an NLOS condition, but only if we know the distance. A similar problem can be observed for $P_{RSS}$. These parameters are not preferred for positioning, but would be useful for other applications such as obstacle detection between two objects with a known inter-object distance. Therefore, $\tau_{RT}$ is a unique parameter which is not correlated with the distance and provides relatively good information about NLOS conditions. More precisely, large values ($\tau_{RT}>20$ ns) imply NLOS conditions, but small values lead to ambiguities. In principle, that means that a probabilistic approach should be used, instead of setting a hard threshold.

Regarding NLOS error mitigation (Fig. \ref{fig:error-mitig}), none of the parameters can perfectly remove the NLOS error, but ${\tau_{MAX}}$ could be chosen since it is slightly better than the rest. The most notable problem is that there are two clouds of samples (corresponding to an error of  around 4 m and 10 m, respectively), associated with the same value of the corresponding parameter (the same problem also appears for other parameters). We also note that larger ${\tau _{RMS}}$ and ${\tau_{MAX}}$ lead to a lower NLOS error, which is consistent with some of the results for tunnels (e.g., \cite{Nerguizian2005}), but stands in contrast to results in indoor environments (e.g.,\cite{Denis2003}). This can be explained by the fact that the tunnel is not a very reflective environment, so for large distances (when the NLOS error is also large) many multi-path components will not be detected.

To summarize, along with ${\tau _1}$ chosen for distance estimation, we choose ${\tau _{RT}}$ for NLOS identification, and ${\tau_{MAX}}$ for NLOS error mitigation. Note that additional refinement (probably, very small) is possible with more than three parameters, but in that case the complexity and communication cost would increase as well.

\subsection{Statistical model for ranging}\label{subsec:stat-models}

\begin{figure}[!tb]
\centerline{
\subfloat[]{\includegraphics[width=0.51\columnwidth]{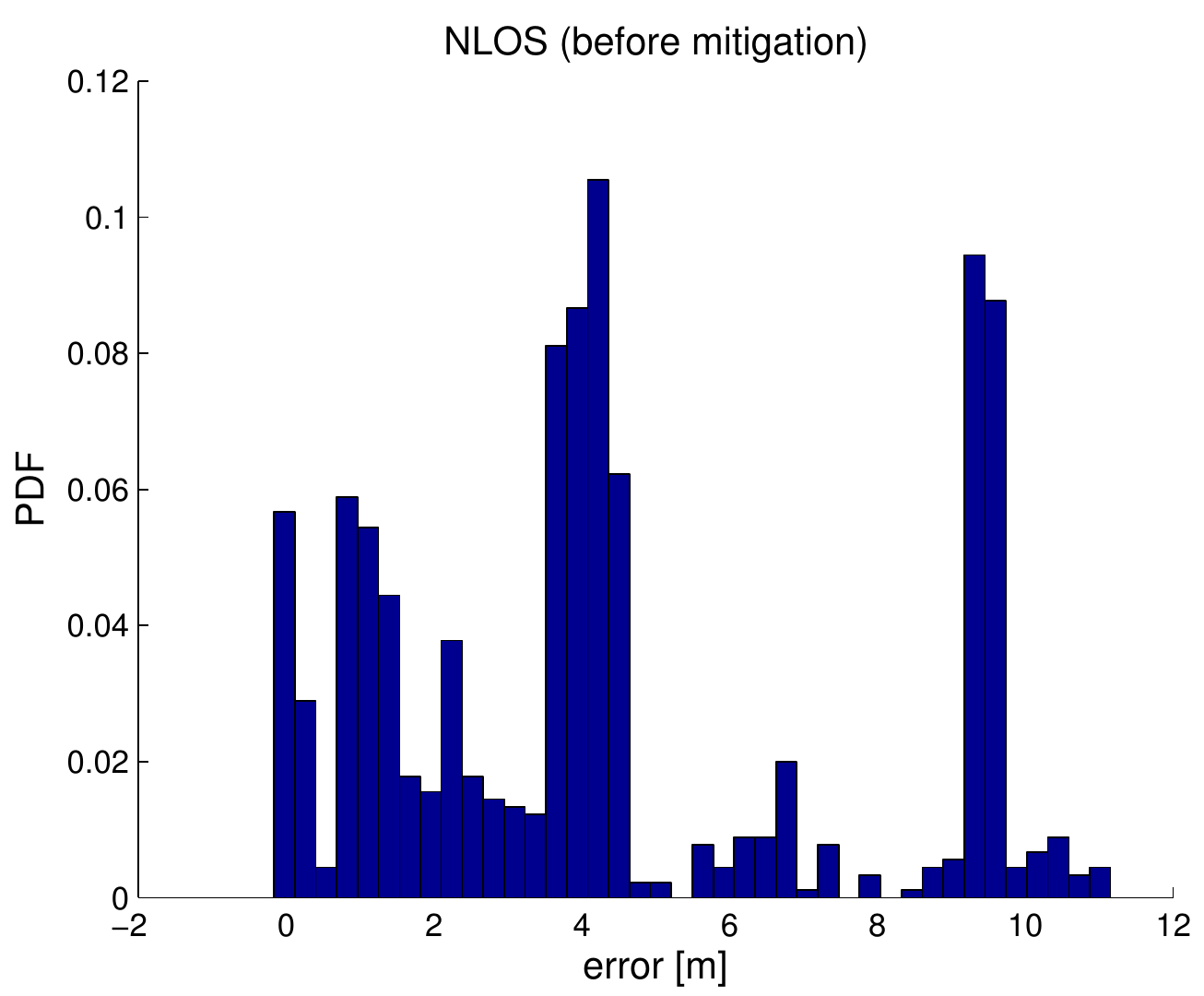}\label{fig:errorTOAnlosBef}}
\hfill
\subfloat[]{\includegraphics[width=0.51\columnwidth]{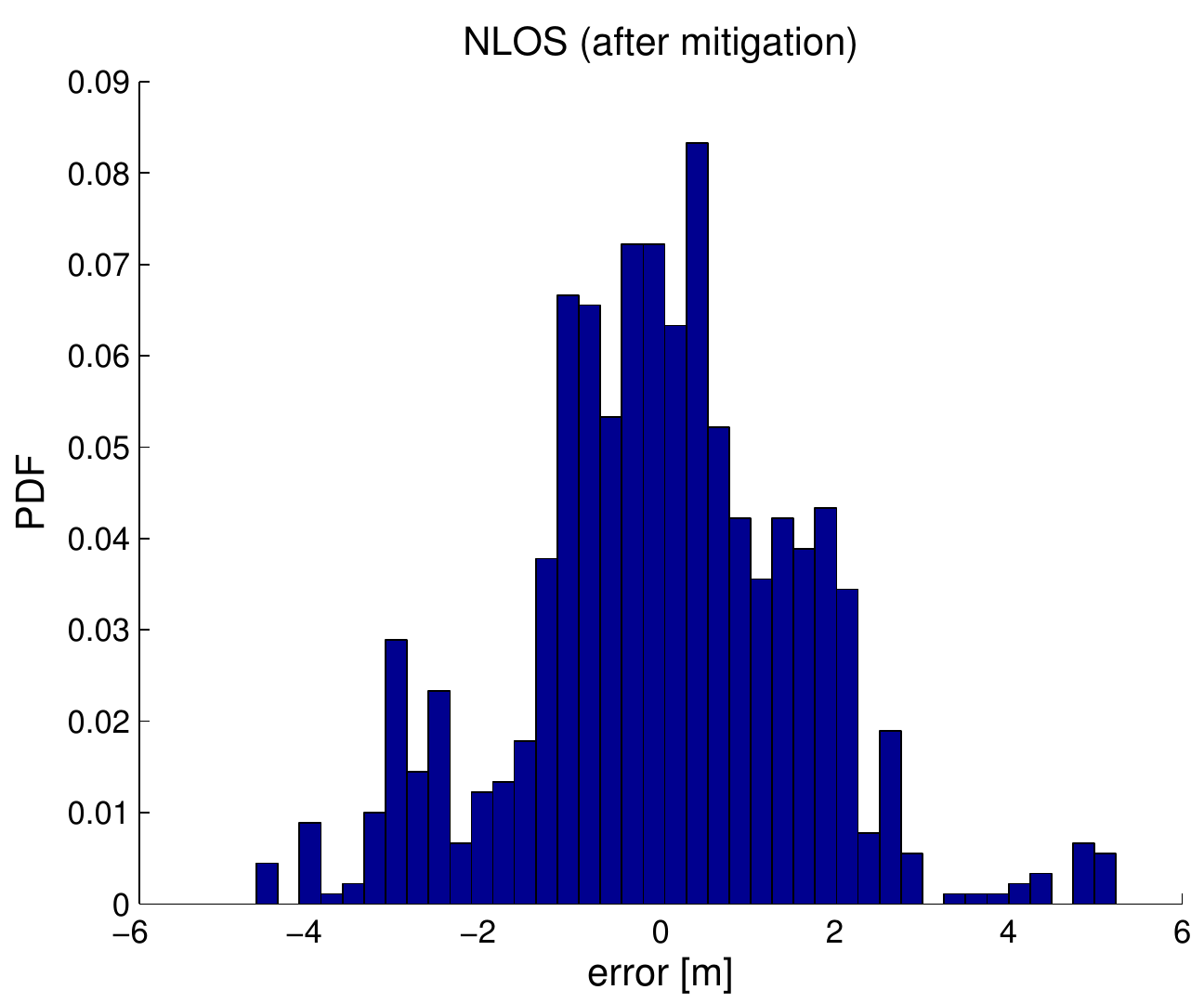}\label{fig:errorTOAnlosAft}}
}
\caption{NLOS error histogram  (a) before, and (b) after error mitigation. The error is not anymore positive (the mean is shifted to zero), and standard deviation is reduced from 3.21 m to 1.61 m.}
\label{fig:errors-nlos}
\end{figure}

Taking into account previous results, we consider the following model:
\be\label{eq:toa-bias}
c{\tau _1} = \left\{ \begin{array}{l}
d + {\mu _L } + {\nu' _L},\,\,\,\,\,\,\,\,\,\,\,\,\,\,\,\,\,\,\,\,{\rm{if}}\,H = {\rm{LOS}}\\
d + g({\tau _{MAX}}) + {\nu'_N},\,\,\,{\rm{if}}\,H = {\rm{NLOS}}
\end{array} \right.
\ee
where ${\nu' _L}$ and ${\nu' _N}$ are noise components, ${\mu _L }=-0.27$ m is a known LOS bias caused by finite bandwidth and false alarms\footnote{Note also that we were not capable to measure precisely the true distances during the experiments, so we introduced an additional error up to 10 cm.} (see also Fig. \ref{fig:toa-dist}d), and $g({\tau _{MAX}})$ is NLOS error. $g(\cdot)$ is found by fitting the samples to second-order polynomial curve, which provides the best fit out of many analyzed curves. Therefore, we set $g({\tau _{MAX}})=p_2{\tau^2_{MAX}}+p_1{\tau _{MAX}}+p_0$. As we can see in Fig. \ref{fig:errors-nlos}, the NLOS error after mitigation is significantly decreased, and its PDF is closer to Gaussian. Taking this result into account (and also Fig. \ref{fig:toa-dist}d), a reasonable model is that ${\nu' _L}$ and ${\nu' _N}$ follow (approximately) zero-mean Gaussian distribution, i.e., ${\nu' _L} \sim \mathcal{N}(0,\sigma_{L}^2)$ and ${\nu' _N} \sim \mathcal{N}(0,\sigma_{N}^2)$.

Finally, in order to estimate $H$, we use Bayes' rule:
\be\label{eq:nlos-bayes}
p(H|{\tau _{RT}}) = \frac{{p({\tau _{RT}}|H)p(H)}}{{\sum_ {H \in \{\rm{LOS}, \rm{NLOS}\}}p({\tau _{RT}}|H)p(H)}}
\ee
where $p(H)$ is the prior, and $p({\tau _{RT}}|H)$ is the likelihood function. Assuming that the floor plan and the detection range are available, a reasonable choice of prior is:
\be\label{eq:prior-rt}
p(H) = \left\{ \begin{array}{l}
1-{S_{N,i}}/{S_{T,i}},\,\,\,{\rm{if}}\,H = {\rm{LOS}}\\
{S_{N,i}}/{S_{T,i}},\,\,\,\,\,\,\,\,\,\,\,\,\,{\rm{if}}\,H = {\rm{NLOS}}
\end{array} \right.
\ee
where $S_{T,i}$ is the total detection area of transmitter $i$, and $S_{N,i}$ is the part of the area corresponding to NLOS caused by tunnel walls.

The likelihood function $p({\tau _{RT}}|H)$ is approximated with an exponential distribution (see Fig. \ref{fig:rtHistAll}), i.e.,
\be\label{eq:lhood-rt}
p({\tau _{RT}}|H) = \left\{ \begin{array}{l}
{\lambda _L}{e^{ - {\lambda _L}{\tau _{RT}}}},\,\,\,\,{\rm{if}}\,H = {\rm{LOS}}\\
{\lambda _N}{e^{ - {\lambda _N}{\tau _{RT}}}},\,\,\,{\rm{if}}\,H = {\rm{NLOS}}
\end{array} \right.
\ee
where the parameters ${\lambda _L}$ and ${\lambda _N}$ are found as the inverse of the sample means (from LOS and NLOS samples, respectively). All parameters are summarized in Table \ref{table:final-model}.

\begin{table}[!tb]
\caption{Estimated parameters for ranging.}
\label{table:final-model}
\centering
\begin{tabular}{c||c}
 Parameters &  Estimated values \\
 \hline\hline
 $[p_2~p_1~p_0]$ & [0.00087 -0.2 11.72] \\\hline
$\sigma_{L}$ & 0.16 m \\\hline
$\sigma_{N}$ & 1.61 m \\\hline
$\lambda_{L}$ & 0.333 ns$^{-1}$ \\\hline
$\lambda_{N}$ & 0.075 ns$^{-1}$\\\hline
\end{tabular}
\end{table}

\begin{figure}[!tb]
\centerline{
\includegraphics[width=0.85\columnwidth]{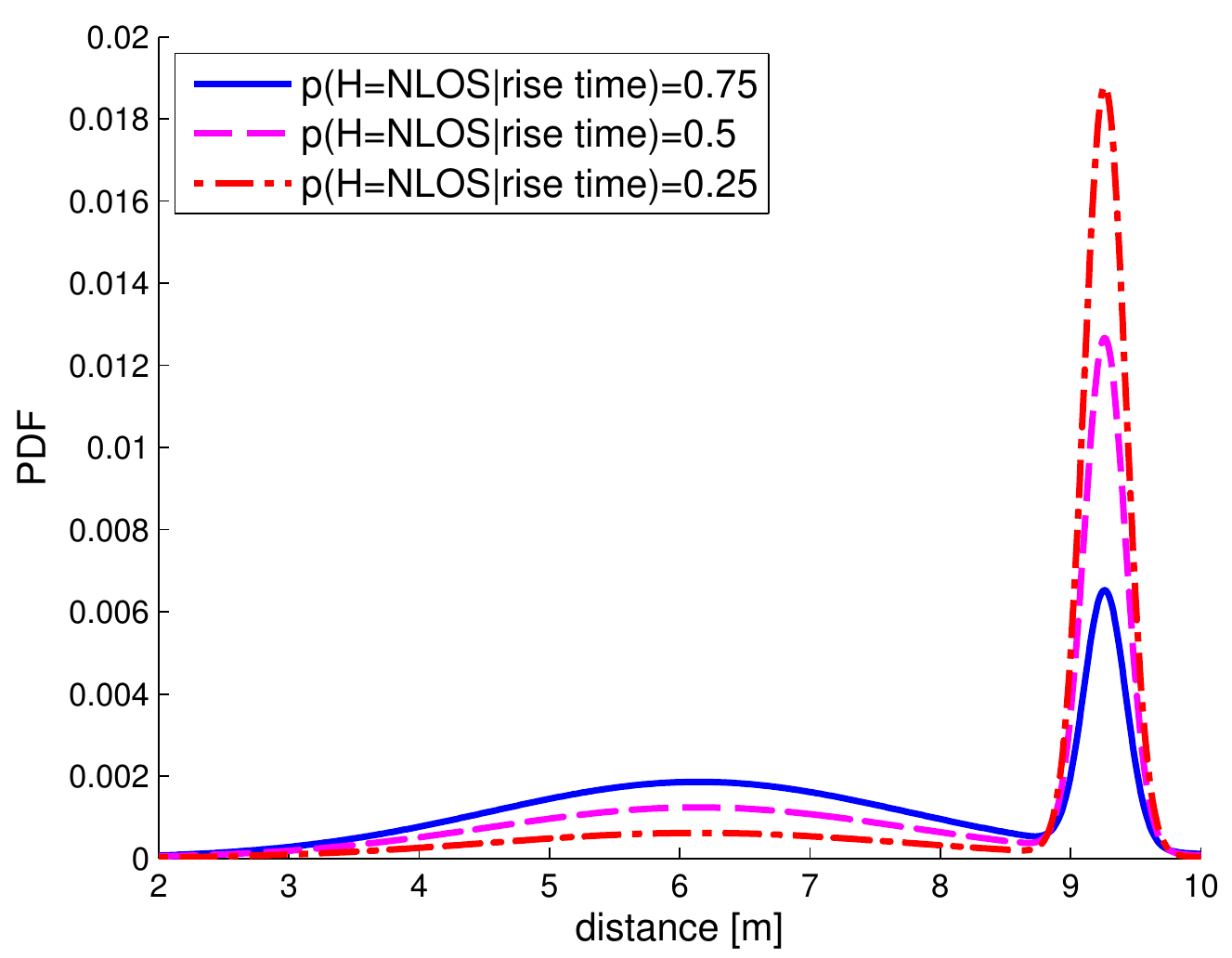}
}
\caption{Likelihood function for $\tau_1=30$ ns, $\tau_{MAX}=60$ ns, and different values of posterior $p(H = {\rm{NLOS}}|{\tau _{RT}})$.}
\label{fig:lhoodDist}
\end{figure}

Traditional (deterministic) positioning techniques (see \cite{Guvenc2009} and references therein) require to make decision on $H$, (e.g., $\hat H=\arg\max_H p(H|{\tau _{RT}})$), and to obtain a point-estimate of the distance as:
\be\label{eq:dist-point}
\hat d = \left\{ \begin{array}{l}
c{\tau _1} - {\mu _L },\,\,\,\,\,\,\,\,\,\,\,\,\,\,\,\,\,\,\,{\rm{if}}\,\,\hat H = {\rm{LOS}}\,\,\\
c{\tau _1} - g({\tau _{MAX}}),\,\,\,{\rm{if}}\,\,\hat H = {\rm{NLOS}}
\end{array} \right.
\ee
\begin{figure}[!tb]
\centerline{
\includegraphics[width=0.85\columnwidth]{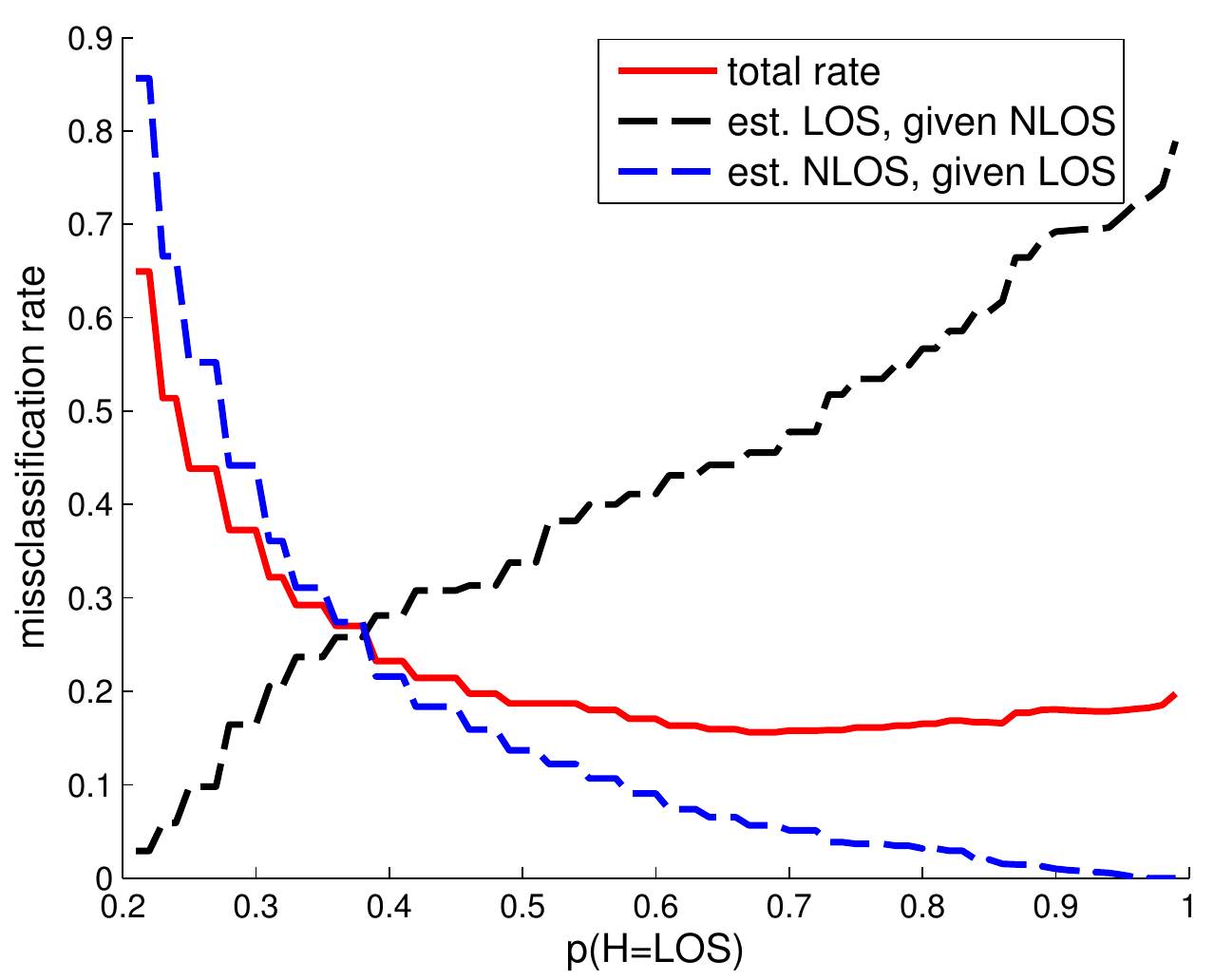}
}
\caption{Misclassification rates estimated from our (LOS and NLOS) sample set. We note that misclassification is unavoidable (especially for small values of $\tau_{RT}$), and that it is highly dependent on prior.}
\label{fig:missRatePrior}
\end{figure}
However, this approach is highly unrecommended due to the high occurrence of misclassifications (see Fig. \ref{fig:rtDistAll} and Fig. \ref{fig:missRatePrior}), and would not provide a full statistical information about the distance. Therefore, it is much better to provide a likelihood function, which takes into account both hypotheses about the LOS/NLOS condition, i.e.,
\beqa\label{eq:dist-lhood}
&p({\bf{e}}|d) = p(H = {\rm{LOS}}|{\tau _{RT}}){\cal N}(c{\tau _1} - {\mu _L } - d;0,\sigma _L^2) + \nonumber \\ 
& p(H = {\rm{NLOS}}|{\tau _{RT}}){\cal N}(c{\tau _1} - g({\tau _{MAX}}) - d;0,\sigma _N^2)
\eeqa
where the vector ${\bf{e}}$ is the set of all available measurements ($\tau_1$, $\tau _{RT}$ and $\tau_{MAX}$). The illustration is shown in Fig. \ref{fig:lhoodDist}. This non-Gaussian likelihood can be used for soft-decision (typically, Bayesian) positioning algorithms. A reasonable option is the algorithm in \cite{Cong04}, which can provide multiple location estimates from the multi-modal PDF. The detailed analysis of positioning algorithms is beyond the scope of this paper.

\section{Comparison with measurements from mine tunnel}\label{sec:kiruna}

\begin{figure}[!tb]
\centerline{
\subfloat[]{\includegraphics[width=0.85\columnwidth]{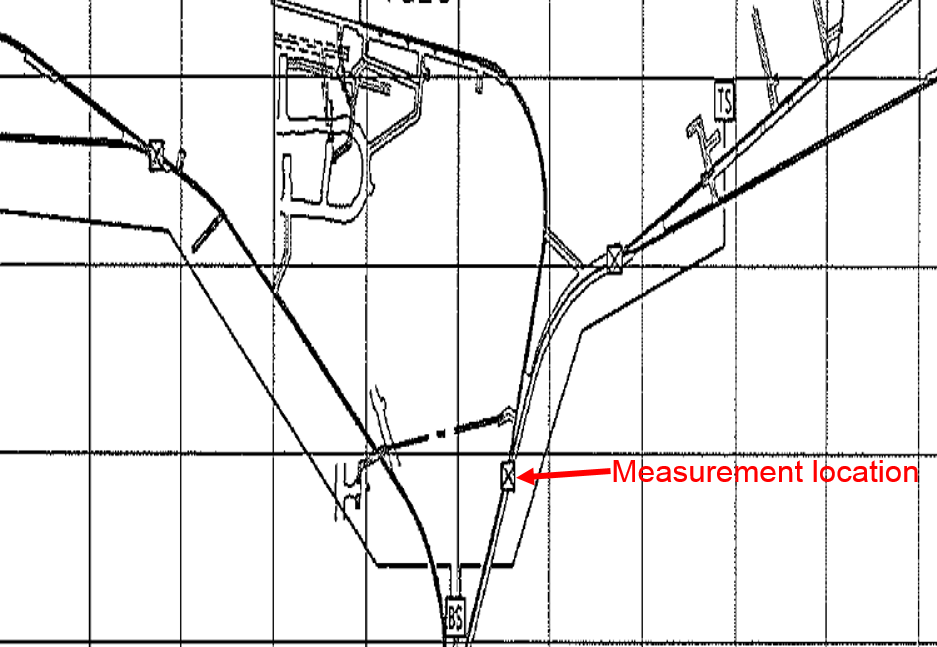}}
}

\centerline{
\subfloat[]{\includegraphics[width=0.85\columnwidth]{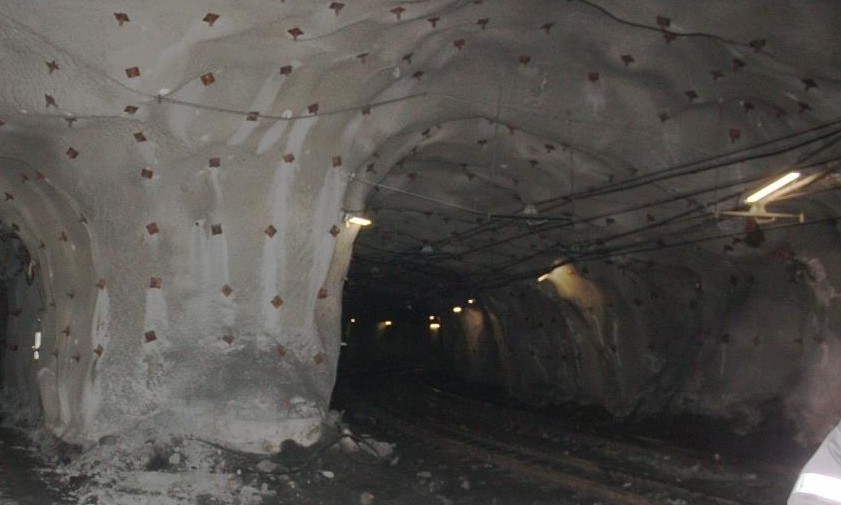}}
}
\caption{(a) A map of the tunnels in Kiruna (located 1045 m below the ground level), and (b) a photo of the measurement location.}
\label{fig:kiruna-wide-tunnel}
\end{figure}

We compare the previous results with the measurements\footnote{This measurement campaign has been performed as a part of another project \cite{Ferrer-Coll2012b}, not directly related with the ranging and positioning. Therefore, there are not enough samples to make the statistical models as in Section \ref{sec:analysis-model}, but these measurements may be useful to give an insight if there is at least a similar behavior as in the LiU tunnel.} obtained in an iron-ore mine, located in Kiruna, Sweden, 1045 m below the ground level. The same measurement setup has been used (described in Section \ref{sec:setup}), but some of the parameters were different: the power level (set to 10 dBm), the central frequencies (set to 1890 MHZ and 2450 MHz), and the bandwidth (set to 500 MHz). Two scenarios have been considered, LOS and NLOS caused by a mine wall, in which 80 PDPs are obtained (56 for LOS, and 24 for the NLOS scenario). The considered tunnel is a semi-arched tunnel in which the walls and ceiling are made of reinforced concrete. The tunnel is 7.1 m wide and 4.67 m high. To measure the NLOS scenario, one antenna was placed in a joint area, where the tunnel is connected with other narrower tunnel. This joint area is 15 m wide and 5.3 m high. A map and a photo of the measurement location are shown in Fig. \ref{fig:kiruna-wide-tunnel}, and the comparison with the LiU tunnel is shown in Table \ref{table:kiruna-liu-geom}. As we can see, the Kiruna tunnel system has a significantly larger complexity than the LiU tunnel.

\begin{table}[!tb]
\caption{Comparison of LiU and Kiruna tunnels.}
\label{table:kiruna-liu-geom}
\centering
\begin{tabular}{c||c | c}
 & LiU tunnel &  Kiruna tunnel \\
 \hline\hline
shape & rectangular & semi-arched \\\hline
material & concrete + metal & concrete \\ \hline
surfaces & flat & rough \\\hline
width x height & 2.9 m x 2.8 m & 7.1 m x 4.67 m\\\hline

\end{tabular}
\end{table}

\begin{figure*}[!tb]
\centerline{
\subfloat[]{\includegraphics[width=0.32\textwidth]{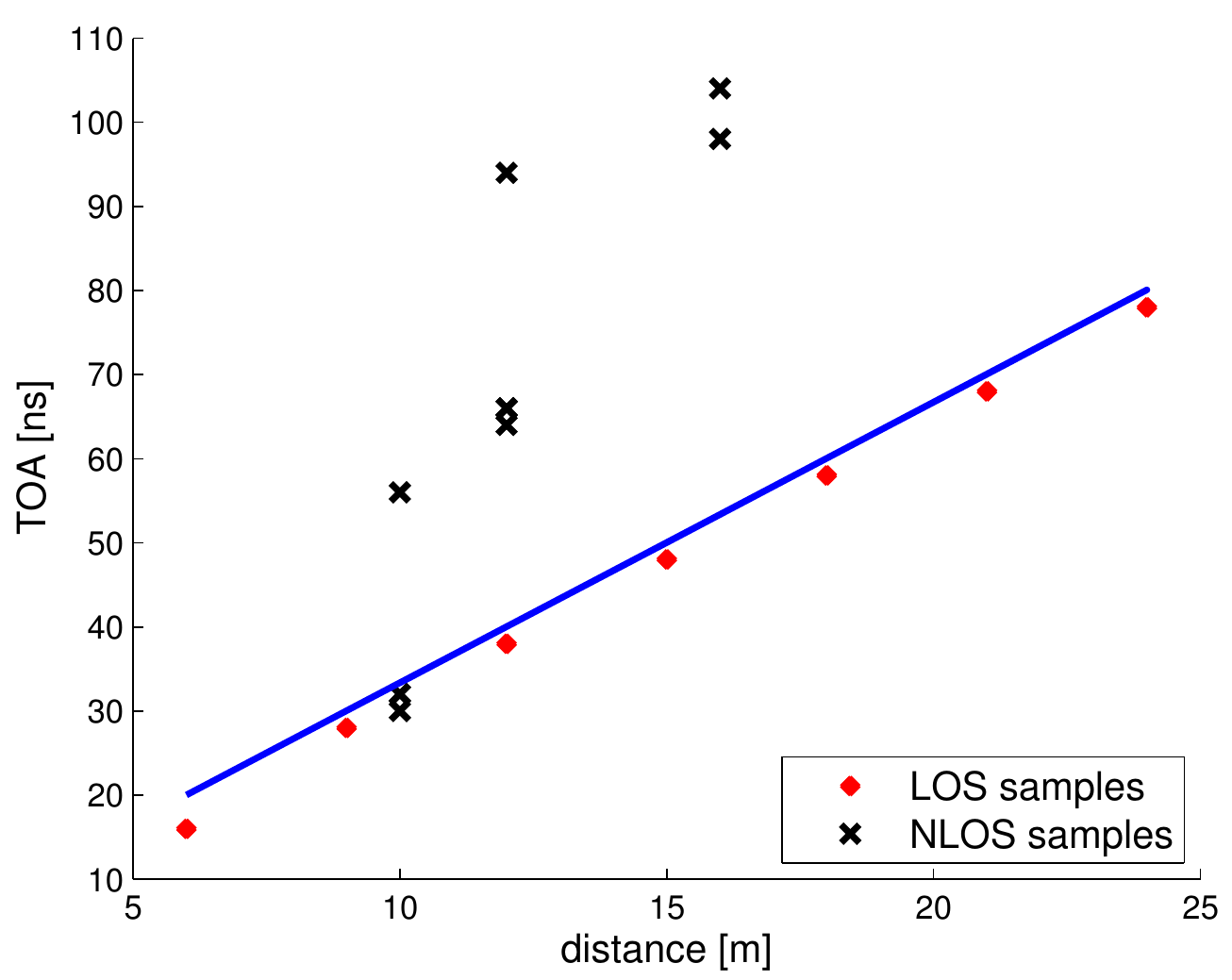}\label{fig:toaVSdistance}}
\hfill
\subfloat[]{\includegraphics[width=0.32\textwidth]{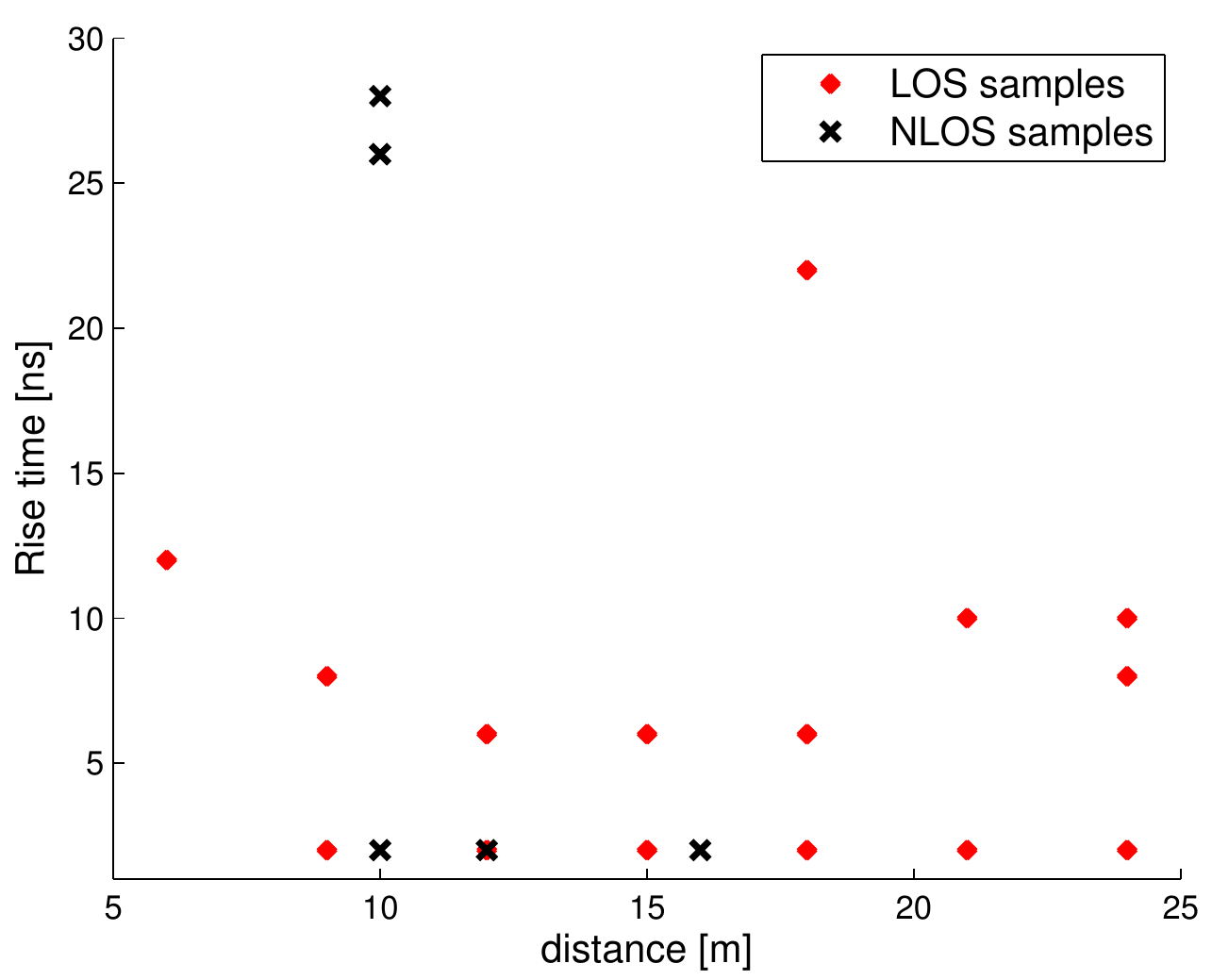}\label{fig:riseTimeVSdistance}}
\hfill
\subfloat[]{\includegraphics[width=0.32\textwidth]{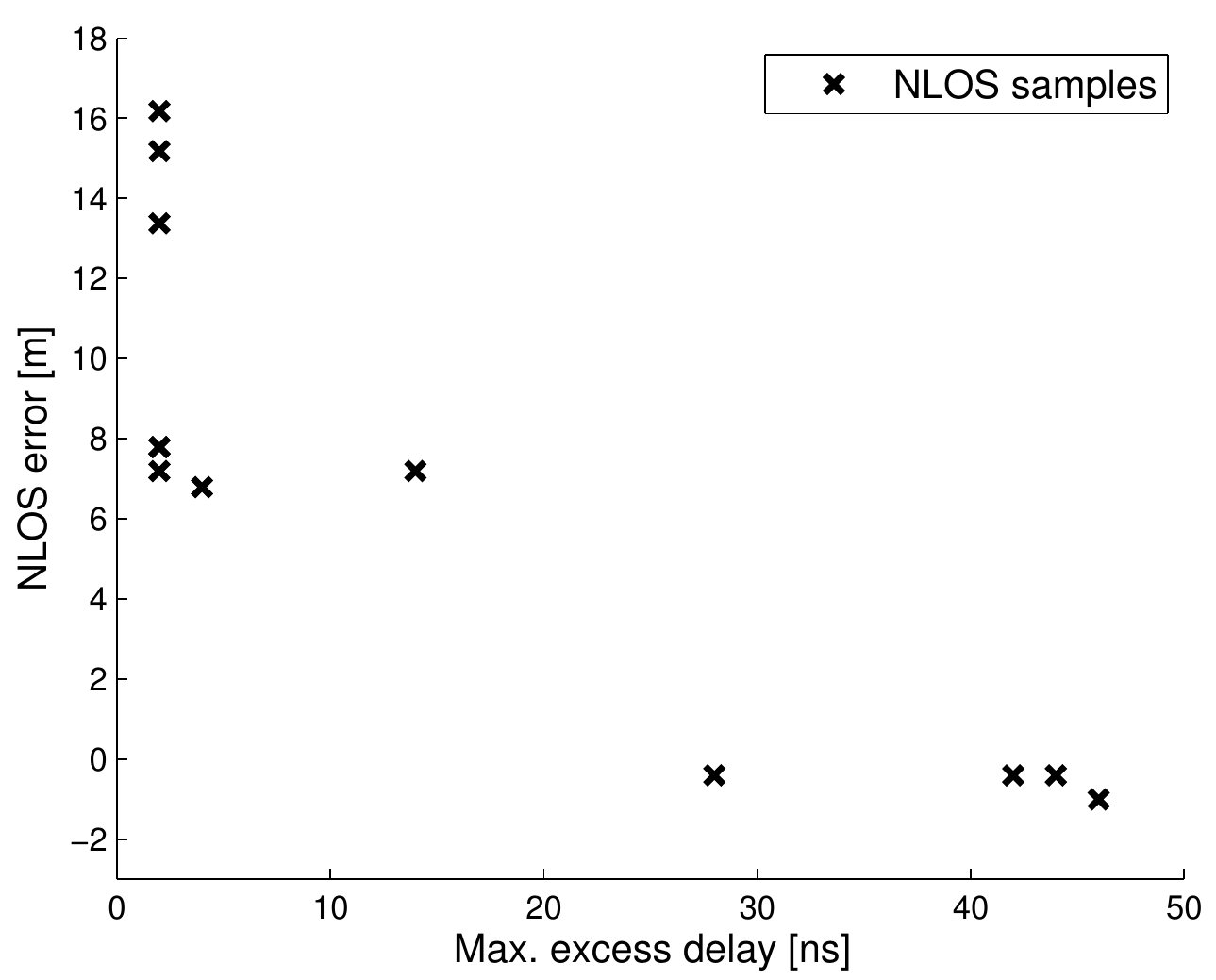}\label{fig:maxExcessVSerror}}
}
\caption{The results from the wide tunnel in Kiruna: (a) TOA vs distance, (b) rise-time vs distance,  and (c) NLOS error vs max. excess delay. These results should be compared with Fig. \ref{fig:distTOAall}, Fig. \ref{fig:rtDistAll} and Fig. \ref{fig:errorMDSnlos}, respectively. Some of the samples overlap since the channel was static during the experiments.}
\label{fig:kiruna-results}
\end{figure*}

The main results are shown in Fig. \ref{fig:kiruna-results}. As in the LiU tunnel, the TOA can provide (Fig. \ref{fig:kiruna-results}a) very accurate distance estimates in the LOS, but there is a large error (up to 16 m) in the NLOS scenario. Therefore, NLOS identification and error mitigation is required for accurate ranging. For that purpose, $\tau_{RT}$ and $\tau_{MAX}$ may again provide some useful information. Large values of $\tau_{RT}$ may imply NLOS condition (Fig. \ref{fig:kiruna-results}b), but there are more ambiguities as compared to the LiU tunnel. In principle, that means that the likelihood (eq. \eqref{eq:lhood-rt}) may be non-informative, so NLOS identification will be based only on the prior (eq. \eqref{eq:prior-rt}). Regarding NLOS error mitigation (Fig. \ref{fig:kiruna-results}c), $\tau_{MAX}$ provides relatively good information about the NLOS error. Moreover, as in the LiU tunnel, larger $\tau_{MAX}$ leads to a lower NLOS error.

\begin{table}[!tb]
\caption{Comparison of main channel propagation parameters.}
\label{table:kiruna-liu-comp}
\centering
\begin{tabular}{c||c | c}
 & LiU tunnel &  Kiruna tunnel \\
 \hline\hline
$\tau _{{\rm{RMS}}}$ [ns] (LOS) & 2.7-15.5 & 2.3-12 \\ 
~~$\tau _{{\rm{RMS}}}$ [ns] (NLOS) & 3.2-16.5 & 0.9-14.5 \\\hline
$\tau _{{\rm{MAX}}}$ [ns] (LOS) & 15-170 & 10-58 \\ 
~~$\tau _{{\rm{MAX}}}$ [ns] (NLOS) & 13-120 & 2-46 \\\hline
$\tau_{{\rm{RT}}}$ [ns] (LOS) & 0.5-19 & 2-22 \\ 
~~$\tau_{{\rm{RT}}}$ [ns] (NLOS) & 0.5-45 & 2-28 \\\hline
$\kappa$ (LOS) & 1-45 & 0.98-4.5 \\ 
~~$\kappa$ (NLOS) & 1-41 & 0.1-5.6 \\\hline
\end{tabular}
\end{table}

We also compare the main channel propagation parameters, shown in Table \ref{table:kiruna-liu-comp}. We note that the values of RMS delay spread are relatively low in both environments, as expected for semi-reflective environments such as tunnels. Obtained max. excess delay values are larger in the LiU tunnel, but this is probably caused by different power levels of the transmitter and the tunnel dimensions. Regarding the rise-time, the values are similar in both environments for LOS, but for the NLOS scenario, a higher rise-time is obtained in the LiU tunnel (which facilitates NLOS identification). Since the rise-time is significant in most cases, the ranging techniques based on the strongest path detection \cite{Dardari2009} are not appropriate in the tunnel environments. Finally, we notice that the values of kurtosis are much lower in Kiruna tunnel, which means that the amplitude of the impulse response can be approximated with a Gaussian distribution (which has kurtosis equal to 3).

In summary, since our measurements have been done for two tunnel geometries, we expect that our results will be relevant for a number of realistic tunnels. Our most general conclusion is that i) TOA-based ranging leads to a large positive bias in case of NLOS, and that ii) a reliable hard decision on whether we have a NLOS condition cannot be made from the channel propagation parameters. The rise time and the maximum excess delay are expected to provide very useful information for NLOS identification and error mitigation, respectively, but it is advisable to reconfirm this result for each particular environment (especially, in case of presence of heavy machinery and vehicles that are not considered in our study). Finally, the parameters of the model (Table \ref{table:final-model}) cannot be generalized and should be re-estimated for each particular environment.

\section{Conclusions and Future Work}\label{sec:conc}
We presented the results of a UWB measurement campaign performed in a basement tunnel of LiU. More specifically, we analyzed channel propagation parameters, selected a subset of parameters that allow relatively accurate TOA-based ranging, and provided an appropriate statistical model. One main result is that the rise-time should be used for NLOS identification, and the maximum excess delay for NLOS error mitigation. The main problem is that an NLOS condition cannot be perfectly determined, so the distance likelihood has to be represented in a Gaussian mixture form. That means that soft-decision algorithms are required for accurate ranging and positioning in tunnels. Finally, taking into account our comparison with the measurements from a mine tunnel, we believe that our main results will be useful for many other tunnel environments. For the future work, we intend to improve ranging and positioning in this kind of environments, especially using non-parametric machine learning techniques. In addition, we plan to develop infrastructure-free cooperative localization algorithms, which are crucial for search-and-rescue operations in GPS-denied environments.


\footnotesize
\bibliographystyle{ieeetr} 
\bibliography{paper-liu-tunnel}

\begin{IEEEbiography} [{\includegraphics[width=1in,height=1.25in,clip,keepaspectratio]{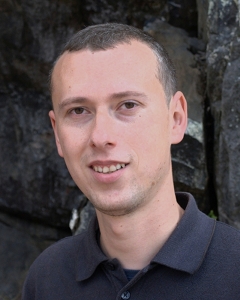}}] {Vladimir Savic}
received the Dipl.Ing. degree in electrical engineering from the University of Belgrade, Belgrade, Serbia, in 2006, and the M.Sc.
and Ph.D. degrees in communications technologies and systems from the Universidad Politecnica de Madrid, Madrid, Spain, in 2009 and 2012, respectively. He was a Digital IC Design Engineer with Elsys Eastern Europe, Belgrade, from 2006 to 2008. From 2008 to 2012, he was a Research Assistant with the Signal Processing Applications Group, Universidad Politecnica de Madrid. He spent three months as a Visiting Researcher at the Stony Brook University, NY, and four months at the Chalmers University of Technology, Gothenburg, Sweden. In 2012, he joined the Communication Systems (CommSys) Division, Link\"{o}ping University, Link\"oping, Sweden, as a Postdoctoral Researcher. He is co-author of more than 25 research papers in the areas of statistical signal processing and wireless communications. His research interests include localization and tracking, wireless channel modeling, Bayesian inference, and distributed and cooperative inference in wireless networks.

\end{IEEEbiography}

\begin{IEEEbiography} [{\includegraphics[width=1in,height=1.25in,clip,keepaspectratio]{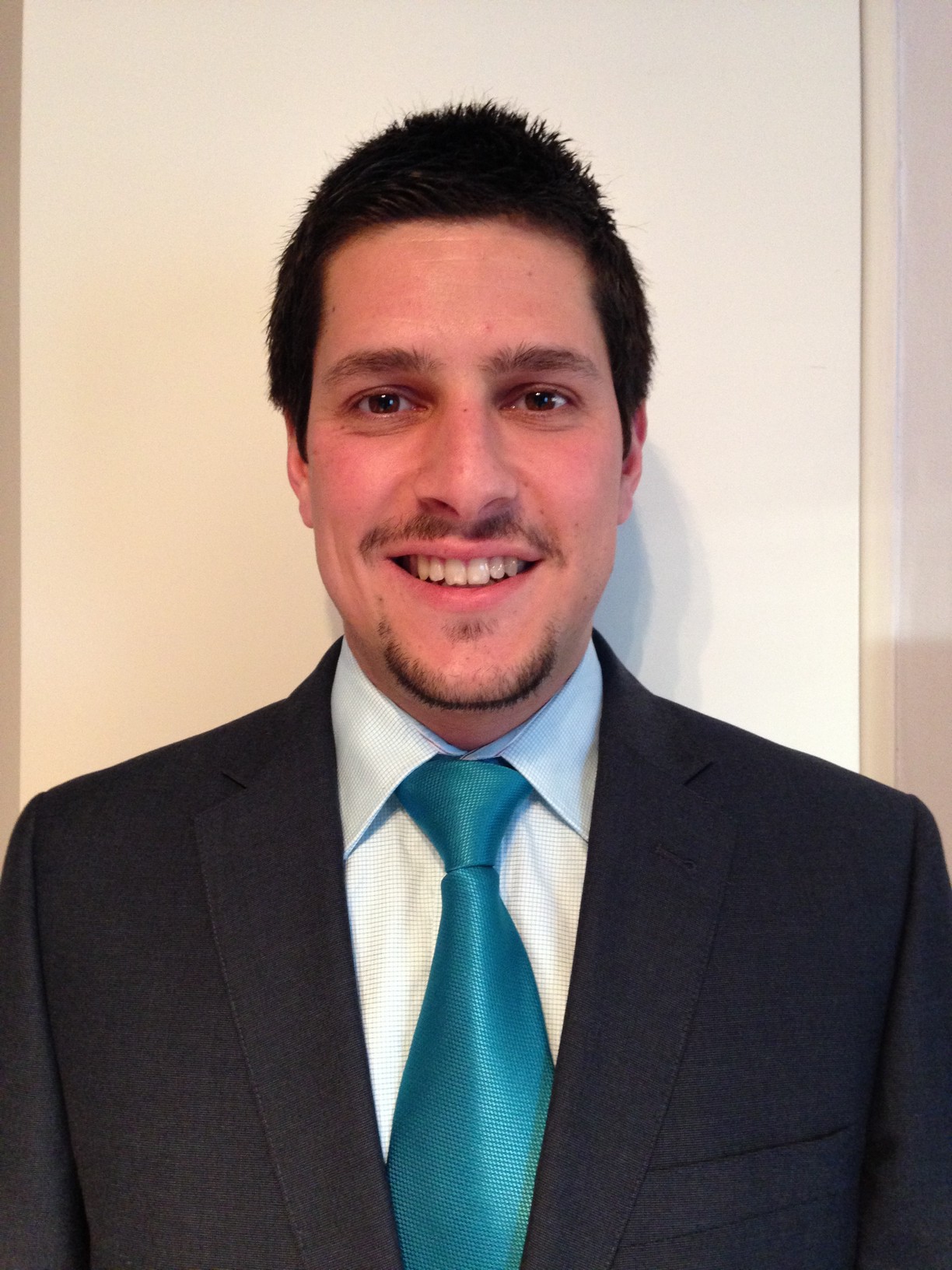}}] {Javier Ferrer-Coll}
received the B.Sc. degree in telecommunication engineering in 2005 and the M.Sc. in 2008 from the Universidad Politecnica de Valencia, Spain. In 2009, he was employed at the university of G\"avle, Sweden. He received his Ph.D in 2014, from the School of Information and Communication, Royal Institute of Technology (KTH), Stockholm, Sweden. He is currently employed as system engineer in the communication department at Combitech AB. His research interest is channel characterization in industrial environments. Specially, measurement system design to extract the channel characteristics, as well as techniques to detect and sup​press electromagnetic interferences.
\end{IEEEbiography}

\begin{IEEEbiography} [{\includegraphics[width=1in,height=1.25in,clip,keepaspectratio]{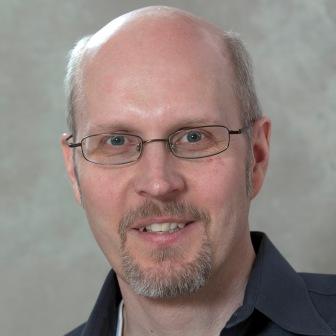}}] {Per \"{A}ngskog}
graduated as an electronics engineer at the University of G\"avle, Sweden, in 1987. Between 1987 and 2000 he
worked in Ericsson AB. Until 1990 he conducted pre-studies of Digital Radio Frequency Memories (DRFM) for airborne
electronic countermeasures. From 1990 to 2000 he worked with design of measurement systems for testing of
transmitters and receivers aimed for mobile telephony systems. Since 2000 he has been employed by the University
of G\"avle, where he currently teaches in the Master’s program in telecommunications and occasionally gives courses
in RF to external companies. Since 2007 he has also been a research engineer in research projects at the Center for RF Measurement Technology, University of G\"avle.
\end{IEEEbiography}

\begin{IEEEbiography} [{\includegraphics[width=1in,height=1.25in,clip,keepaspectratio]{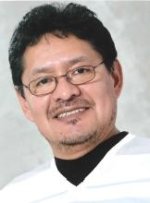}}] {Jos\'{e} Chilo}
has got his B.Sc. degree in industrial engineering from University of San Agustin Arequipa, Per\'u in 1986, and B.Sc. degree in electrical engineering from Royal Institute of Technology (KTH), Kista, Sweden in 2001. From 1986 until he began his studies at the Royal Institute of Technology in 1998, he worked as a consultant in various industries such as textiles, garments manufacturing, transportation, etc., both in Per\'u and Sweden. He has got his PhD in physics from the Royal Institute of Technology in 2008 where he conducted research on signal and data processing, in particular advanced measuring techniques and event classification. He is presently Associate Professor and University Senior Lecturer at University of G\"avle in Sweden.
\end{IEEEbiography}

\begin{IEEEbiography} [{\includegraphics[width=1in,height=1.25in,clip,keepaspectratio]{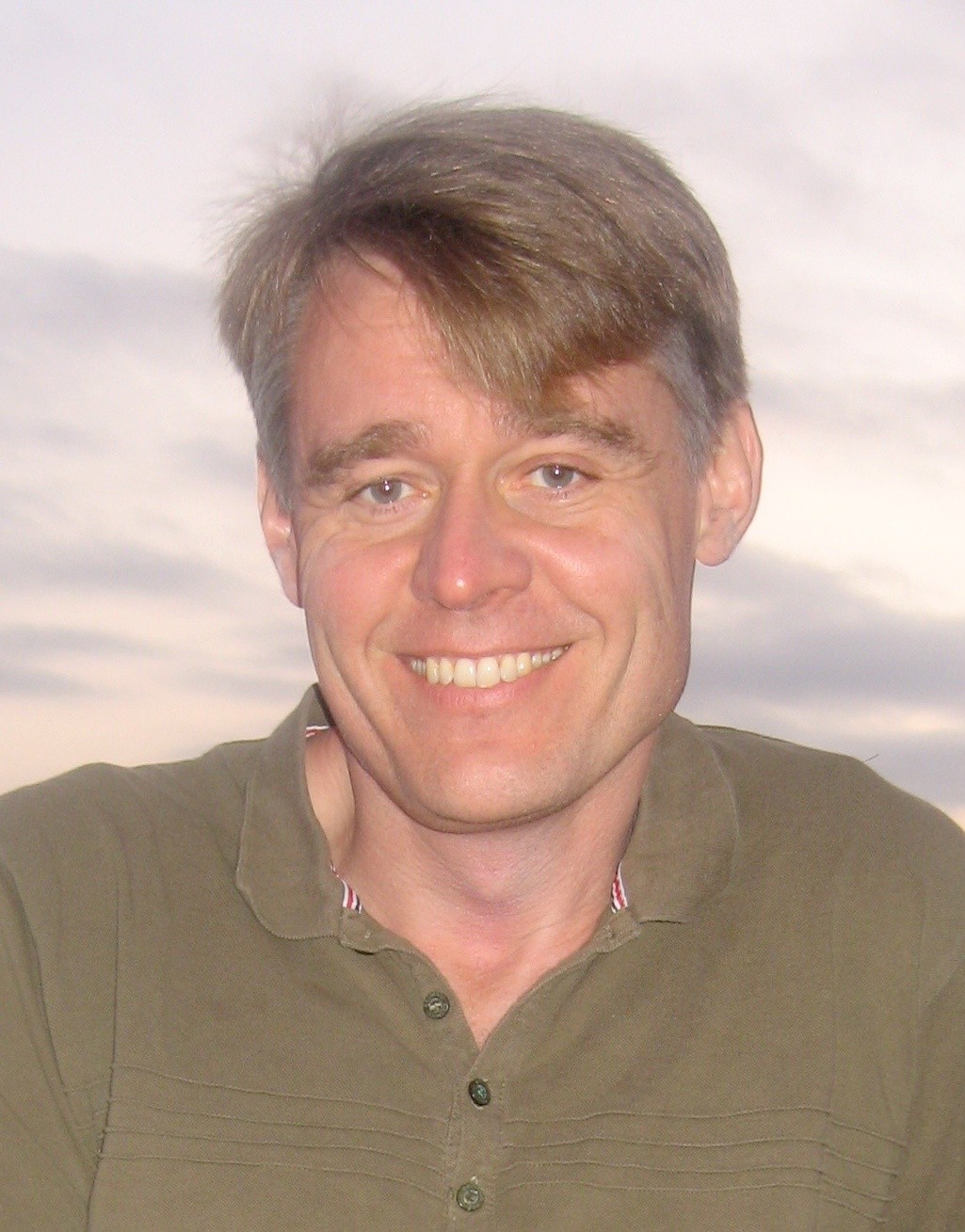}}] {Peter Stenumgaard}
is currently the Research Director in robust wireless communications at the Swedish Defence Research Agency (FOI) in Link\"oping, Sweden. He received the PhD degree in radio communications from the Royal Institute of Technology, Stockholm in 2001. Until 1995 he was a systems development engineer at Saab Military Aircraft, Link\"oping, where he worked with electromagnetic compatibility issues in aircraft design. During 2006 – 2012 he was an adjunct professor at the Center for Radio Measurement Technology at the University of G\"avle, Sweden. He was an adjunct Professor at the Division for Communication Systems in the Department of Electrical Engineering (ISY) at Link\"oping University (LiU) in 2011-2013. He was also the Director of the graduate school Forum Securitatis within security \& public safety at Link\"oping University 2010-2014. His research interests are robust telecommunications for military-, security-, safety- and industrial use.
\end{IEEEbiography}

\begin{IEEEbiography} [{\includegraphics[width=1in,height=1.25in,clip,keepaspectratio]{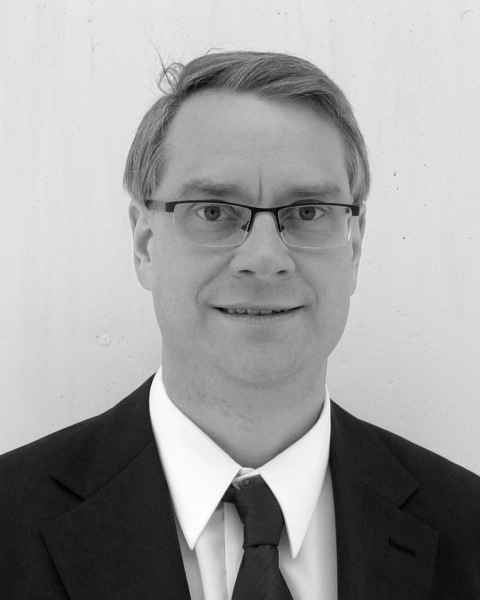}}] {Erik G. Larsson}
received his Ph.D. degree from Uppsala University, Sweden, in 2002.  Since 2007, he is Professor and Head of the Division for Communication Systems in the Department of Electrical Engineering (ISY) at Link\"oping University (LiU) in Link\"oping, Sweden. He has previously been Associate Professor (Docent) at the Royal Institute of Technology (KTH) in Stockholm, Sweden, and Assistant Professor at the University of Florida and the George Washington University, USA.

His main professional interests are within the areas of wireless communications and signal processing. He has published some 100 journal papers on these topics, he is co-author of the textbook \emph{Space-Time Block Coding for Wireless Communications} (Cambridge Univ. Press, 2003) and he holds 10 issued and many pending patents on wireless technology. He is Associate Editor for the \emph{IEEE Transactions on Communications} and he has previously been Associate Editor for several other IEEE journals. He serves as vice-chair for the IEEE Signal Processing Society SPCOM technical committee in 2014. He also serves as chair of the steering committee for the \emph{IEEE Wireless Communications Letters} in 2014--2015.  He is active in conference organization, for example as General Chair of the Asilomar Conference on Signals, Systems and Computers in 2015 (he was Technical Chair in 2012).  He received the 2012 \emph{IEEE Signal Processing Magazine} Best Column Award.
\end{IEEEbiography}

\end{document}